%% file: main_IEEE.tex
\documentclass[compsoc,conference,a4paper,10pt,times]{IEEEtran} 

\usepackage[dvipsnames]{xcolor}

\usepackage[colorlinks=true,linkcolor=blue,breaklinks=True,citecolor=brown,urlcolor=blue]{hyperref}

\usepackage{dingbat}

\usepackage{adjustbox}
\usepackage[linesnumbered,algo2e,ruled]{algorithm2e}
\usepackage{multirow}
\usepackage{booktabs,subcaption,dcolumn}
\usepackage{tikz}
\usepackage{enumitem}
\setlist[itemize]{noitemsep, topsep=0pt}
\usepackage{tabularx}
\usepackage[normalem]{ulem}

\usepackage{pifont}
\usepackage{svg}
\usepackage{soul}
\usepackage[font=footnotesize]{caption}
\usepackage[noadjust]{cite}

\usepackage{amsmath}
\usepackage{amsfonts, amssymb}
\usepackage{colortbl}
\usepackage{tablefootnote}
\usepackage[most]{tcolorbox}
\usepackage[sort&compress, numbers]{natbib}

\AtBeginDocument{%
  \providecommand\BibTeX{{%
    \normalfont B\kern-0.5em{\scshape i\kern-0.25em b}\kern-0.8em\TeX}}}

\input{commands.tex}

\begin{document}

\input{title}

\input{authors.tex}

\pagestyle{plain}
\maketitle

\input{sections/0-abstract}

\input{structure}

\bibliographystyle{IEEEtran}

{\footnotesize
\bibliography{bibliography} 
}

\appendix
\input{appendix/structure_app}

\end{document}

%% file: commands.tex
\newcolumntype{?}{!{\vrule width 1.5pt}}

\newcommand{\textbox}[1]{
    \noindent\fbox{%
        \parbox{0.97\columnwidth}{%
            {#1}
        }%
    }
}

\newtcolorbox{cooltextbox}[1][]{%
    colback=black!5,
    colframe=black!5,
    notitle,
    sharp corners,
    borderline west={0pt}{0pt}{red!80!black},
    enhanced,
    breakable,
    left=0pt,
    right=0pt,
    top=0pt,
    bottom=0pt
    }

\newcommand\revision[1]{%
  \bgroup
  \hskip0pt\color{blue!80!black}%
  #1%
  \egroup
}

%% file: title.tex
\title{The Dark Side of the Web: \\ Towards Understanding Various Data Sources in Cyber Threat Intelligence}






%% file: authors.tex
\author{
\hspace{-0cm}
\IEEEauthorblockN{
Saskia Laura Schröer\IEEEauthorrefmark{1}, Noè Canevascini\IEEEauthorrefmark{2}, Irdin Pekaric\IEEEauthorrefmark{1}, Philine Widmer\IEEEauthorrefmark{3}, Pavel Laskov\IEEEauthorrefmark{1}
\\}
\IEEEauthorblockA{{ 
\IEEEauthorrefmark{1}\textit{University of Liechtenstein},
\IEEEauthorrefmark{2}\textit{ETH Zurich}
\IEEEauthorrefmark{3}\textit{Paris School of Economics}
}
\\
{\small Corresponding authors: saskia.schroeer@uni.li, irdin.pekaric@uni.li, philine.widmer@psemail.eu
}
}}

%% file: sections/0-abstract.tex
\begin{abstract}
\noindent
Cyber threats have become increasingly prevalent and sophisticated. Prior work has extracted actionable cyber threat intelligence (CTI), such as indicators of compromise, tactics, techniques, and procedures (TTPs), or threat feeds from various sources: open source data (e.g., social networks), internal intelligence (e.g., log data), and ``first-hand'' communications from cybercriminals (e.g., underground forums, chats, darknet websites). However, ``first-hand'' data sources remain underutilized because it is difficult to access or scrape their data. 
  
In this work, we analyze (i) 6.6 million posts, 
(ii) 3.4 million messages, 
and (iii) 120,000 darknet websites. We combine NLP tools to address several challenges in analyzing such data. First, even on dedicated platforms, only some content is CTI-relevant, requiring effective filtering. 
Second, ``first-hand'' data can be CTI-relevant from a technical or strategic viewpoint. We demonstrate how to organize content along this distinction. Third, we describe the topics discussed and how ``first-hand'' data sources differ from each other. 
According to our filtering, 20\% of our sample is CTI-relevant. Most of the CTI-relevant data focuses on strategic rather than technical discussions. Credit card-related crime is the most prevalent topic on darknet websites. 
On underground forums and chat channels, account and subscription selling is discussed most. Topic diversity is higher on underground forums and chat channels than on darknet websites. Our analyses suggest that different platforms may be used for activities with varying complexity and risks for criminals. 

\end{abstract}

%% file: structure.tex
\input{sections/1-introduction}

\input{sections/2-related}

\input{sections/3-method}

\input{sections/4-results}

\input{sections/5-discussion}
\input{sections/6-conclusions}

\input{sections/7-ethical}

\input{sections/8-acknowledgment}

%% file: sections/1-introduction.tex
\section{Introduction}
\label{sec:introduction}
\noindent
The Internet offers ample opportunity for cybercriminals and hackers to share information, discuss strategies, and trade illegal goods. This digital underworld is most often accessible through darknet websites, underground forums, and encrypted communication channels and can serve as a breeding ground for cybersecurity threats. A large part of this hidden ecosystem consists of onion sites: websites on the darknet that use the ``.onion'' top-level domain~\cite{takaaki2019dark}. These sites utilize, for instance, The Onion Router (Tor) software or the Invisible Internet Project (I2P) to encrypt their connections, enable anonymous communication, and hide users' identity and location.
Besides darknet websites, malicious actors utilize forums on ``non-onion'' websites. Such forums on the clearnet sometimes present themselves as innocent discussion boards or specialized interest groups. Furthermore, widely used chat applications such as Discord and Telegram have become popular platforms for exchanging information between cybercriminals~\cite{bijmans2021catching}. These mainstream apps allow for end-to-end encryption, self-destructing messages, and creating private channels. Such features make these platforms attractive to cybercriminals as they blur the line between visible and hidden parts of the Internet. Hence, these emerging platforms present new challenges for cybersecurity professionals and law enforcement agencies to monitor and mitigate cyber threats.

Cybercriminals' using these platforms implies that researchers and security professionals can use their content to study anti-social activities, such as private interactions on underground forums~\cite{sun2019understanding}, privacy risks due to unauthorized information flows or identity cloning attacks~\cite{carminati2011probability, jin2011towards}, image data on darknet websites~\cite{jeziorowski2020towards}, or even the spectrum of content found on darknet websites~\cite{takaaki2019dark}.  
In this work, we focus on extracting cyber threat intelligence (CTI) from multiple platforms. The CTI process involves gathering, analyzing, and disseminating information about potential or current attacks that threaten an organization~\cite{saeed2023systematic}. Traditional CTI sources include internal data (such as network logs and incident reports), external threat reports from vendors, various databases, and online social networks like Twitter/X. However, these traditional sources incompletely reflect the evolving threat landscape. For instance, research on existing hacker ecosystems in clearnet data sources, such as the study by Islam et al.~\cite{islam2021hackerscope}, can be enhanced by considering darknet sources.

\begin{figure*}[!htbp]
    \centering
    \includegraphics[width=1\textwidth]{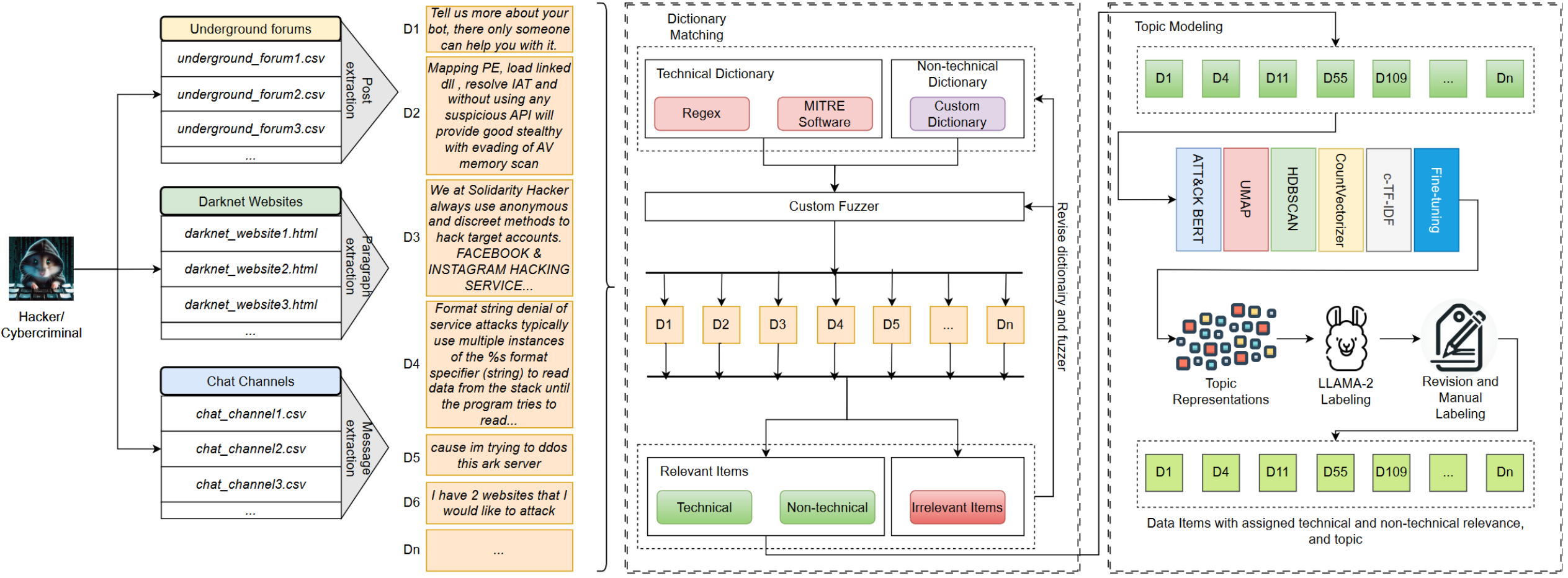}
    \vspace{-0.5cm}
    \caption{\textbf{Our NLP Pipeline.} 
    \textmd{\footnotesize The figure describes our NLP pipeline for a comprehensive analysis of heterogeneous data sources. Relevant versus not relevant data items concern their relevance to CTI. For topic modeling we use BERTopic~\cite{grootendorst2022bertopic}. We describe the details of our pipeline in §\ref{ssec:nlp-pipeline}.}} 
    \label{fig:pipeline}
    \vspace{-3mm}
\end{figure*}

Our work expands the scope of CTI sources to include darknet websites, underground forums, and encrypted communication chat channels (specifically Telegram and Discord), which are comparatively underutilized in prior work (cf. §\ref{sec:related}). Although rich in intelligence, these sources are more difficult for researchers to access: They often require lengthy and complex processes to gain entry~\cite{sakellariou2023secdfan,christin2024learned} and to process them. Additionally, in a recent review of previous studies on darknet content, Basheer and Alkhatib~\cite{basheer2021threats} suggest that CTI would benefit from more natural language processing (NLP).

Previous research has often considered only a single data source (see our literature review in §\ref{sec:related}). We combine diverse hard-to-access data sources and analyze them consistently. First, we classify content into relevant versus not relevant to CTI. Then, we label data items among relevant content as covering technical CTI versus non-technical CTI. This allows both technical and non-technical stakeholders to understand, interpret, and act upon the security information relevant to their roles. Finally, we document the prevalence of different topics across sources. In sum, we address the following research questions: (i) How can we effectively distinguish between relevant and non-relevant CTI content across diverse data sources (RQ1)?; (ii) What content share pertains to technical and non-technical CTI (RQ2)?; and (iii) What type of CTI-related topics can be extracted from darknet websites, underground forums, and Telegram and Discord chat channel data (RQ3)? 
We adapt our NLP tools to accommodate the domain-specific terminology. We highlight potential as well as limitations in using pre-trained neural networks, including large language models, for CTI extraction.

\noindent{\textbf{Contributions.}}
Our primary contribution is a meta-analysis of several ``first-hand'' cyber-crime-related data sources (underground forums, chat channels, and darknet websites).
In addition, we provide our NLP pipeline (from which we derived our results) as an open-source tool. In detail, our work makes several key contributions:
\begin{itemize}
\item We conduct a meta-analysis of multiple ``dark'' data sources. Thereby, we provide guidance on analyzing such sources and highlight differences among ``first-hand'' cybercrime-related data sources.
\item Our meta-analysis builds on our systematic literature review of CTI extraction research (§\ref{sec:related}). It includes work from the past five years (2019--2023), focusing on various sources and NLP methods.
\item We open-source the NLP pipeline (see Figure~\ref{fig:pipeline}) that underlies our meta-analysis. Our pipeline includes specialized cybersecurity dictionaries (see details in §\ref{ssec:nlp-pipeline}) and deep learning methods. By open-sourcing our tools, we foster reproducibility~\cite{repository} and seek to advance the use of NLP and AI in CTI research.
\item We identify and analyze both technical and non-technical CTI in the investigated data sources (§\ref{ssec:technicality}).
\item We present the first comprehensive large-scale analysis of cybercrime-related topics across darknet websites, underground forums, and chat channels (§\ref{ssec:relevance}-§\ref{ssec:topics}). To the best of our knowledge, this is the first work aiming to extract CTI consistently from multiple data sources, and specifically from chat channels.
\end{itemize}


%% file: sections/2-related.tex
\section{Related Work}
\label{sec:related}
\begin{figure}[b]
    \centering
    \includegraphics[width=0.97\columnwidth]{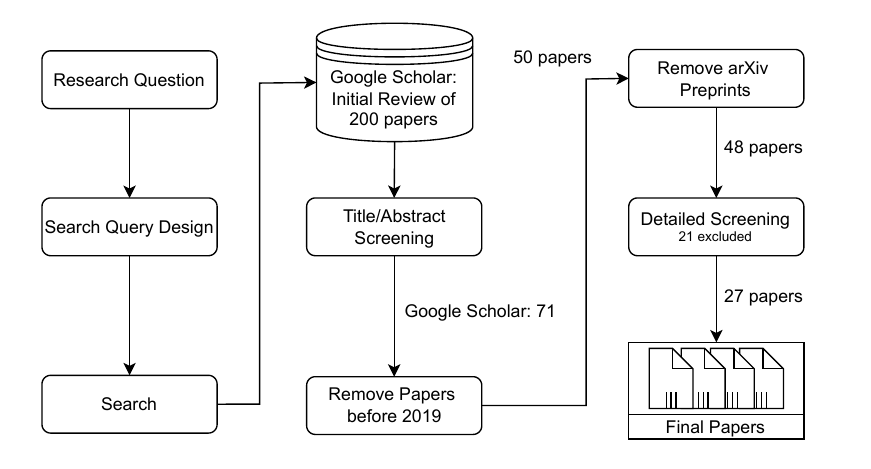}
    \caption{\textbf{Systematic Literature Review.}
    \textmd{\footnotesize The figure describes the systematic literature review conducted in March 2024 to identify the state of the art. We exclude survey papers and literature reviews. }} 
    \label{fig:lit-review}
    \vspace{-3mm}
\end{figure}

\noindent
The field of cybersecurity has seen significant advancements in recent years, particularly in CTI extraction. To document the state of the art, we conduct a systematic literature review,\footnote{The detailed classification of related work can be found in our public repository~\cite{repository}.} examining the most relevant NLP approaches published in the last five years (2019--2023). Our review focuses on Google Scholar and yields 27 relevant papers. We illustrate the detailed steps of the systematic literature review in Figure~\ref{fig:lit-review}. We compare the identified works in the following paragraphs based on (i) the used data sources and (ii) the employed NLP methods.

\begin{figure}[!htbp]
    \centering
    \centering
    \includegraphics[width=1\columnwidth]{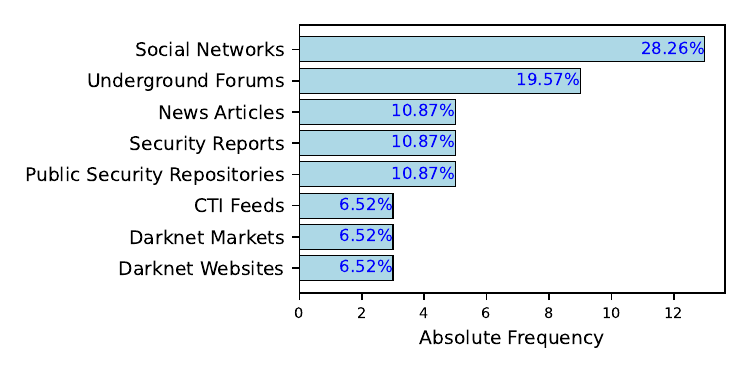}
    \vspace{-5mm}
    \caption{\textbf{Summary of Literature Review: Data Sources.} 
    \textmd{\footnotesize We observe a low number of ``dark'' data sources in the analyzed 27 works. While the number of underground forums seems comparatively high, most of these works examine a single forum rather than multiple ones. Also, we do not identify any paper reviewing chat channels such as Telegram or Discord in the context of CTI extraction. When data sources are used in combination, they are mostly from the clearnet.}} 
    \label{fig:rel_ds}
\end{figure}

\noindent \textbf{Data Sources.} Sauerwein et al.~\cite{sauerwein2019analysis} reveal a diverse range of categories for CTI sources: open-source security data (social networks (SN)~\cite{shin2021twiti, koloveas2021intime, tekin2021obtaining, adewopo2020exploring, zhao2020timiner, kristiansen2020cti, rodriguez2020enhancing, bose2021tracing, purba2023extracting, sufi2023new, sufi2023global, zenebe2019cyber}, news articles (NA)~\cite{zhao2020timiner, rodriguez2020enhancing, zhou2022cti, sufi2023new, panagiotou2021towards}, security reports (SR)~\cite{adewopo2020exploring, zhao2020timiner, rodriguez2020enhancing, zhou2022cti, sun2021automatic}, and public security repositories (SD)~\cite{adewopo2020exploring, ampel2020labeling, purba2023extracting, sufi2023global}), internal intelligence (data generated within organizations including log data and internal feeds\footnote{This data is normally not shared with the public, which may be why none of the related works considers such data.}), and CTI feeds (FE)~\cite{azevedo2019pure, guo2023framework, cha2020blockchain} (intelligence feeds provided by specialized vendors or community-driven platforms). In the realm of open-source security data, there is a high focus on social networks, especially on Twitter (now X), due to its API with historically generous research access. Other open sources are less utilized in comparison. Notably, related work often does not explicitly state the reasons for selecting specific data sources, limiting the understanding of source selection criteria and increasing potential biases. While most of the aforementioned sources have been studied and employed in state-of-the-art approaches, there is a notable underutilization of ``first-hand'' information from cybercriminals. This category includes data from darknet websites (DW)~\cite{zhao2020timiner, furumoto2021extracting} and marketplaces (DM)~\cite{arnold2019dark, ampel2020labeling, furumoto2021extracting}, underground forum discussions (UF)~\cite{adewopo2020exploring, zhao2020timiner, hossen2021generating, kadoguchi2020deep, arnold2019dark, ampel2020labeling, huang2020monitoring, furumoto2021extracting, gautam2020hacker}, and encrypted chat channels. Regarding these ``dark'' data sources (or ``first-hand'' data sources from cybercriminals), the focus is mostly on underground forums, but the data is usually obtained from a single forum rather than multiple different ones. Overall, these ``hard-to-access'' sources are typically analyzed in isolation, limiting comprehensive insights. We visualize our results in Figure~\ref{fig:rel_ds}.

\begin{figure}[!htbp]
    \centering
    \centering
    \includegraphics[width=1\columnwidth]{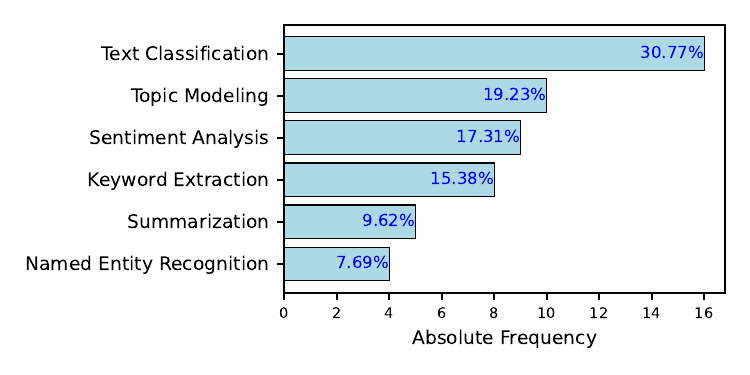}
    \vspace{-5mm}
    \caption{\textbf{Summary of Literature Review: NLP Methods.} 
    \textmd{\footnotesize The main NLP method applied in prior work is Text Classification, followed by Topic Modeling. Please note that some papers use a combination of multiple NLP methods.}} 
    \label{fig:rel_nlp}
\end{figure}

\noindent \textbf{NLP Methods.} NLP methods are popular in CTI extraction. These methods (as outlined by Arazzi et al.~\cite{arazzi2023nlp}) include Named Entity Recognition (NER)~\cite{shin2021twiti, koloveas2021intime, zhou2022cti, panagiotou2021towards}, Topic Modeling (TM)~\cite{kristiansen2020cti, rodriguez2020enhancing, azevedo2019pure, hossen2021generating, bose2021tracing, kadoguchi2020deep, sakellariou2023secdfan, arnold2019dark, ampel2020labeling, sufi2023new}, Sentiment Analysis (SA)~\cite{zhao2020timiner, bose2021tracing, sakellariou2023secdfan, arnold2019dark, tundis2022feature, sufi2023global, sufi2023new, zenebe2019cyber}, Text Classification (TC)~\cite{shin2021twiti, adewopo2020exploring, zhao2020timiner, koloveas2021intime, tekin2021obtaining, kristiansen2020cti, rodriguez2020enhancing, azevedo2019pure, ampel2020labeling, hossen2021generating, purba2023extracting, sun2021automatic, sufi2023new, huang2020monitoring, panagiotou2021towards, gautam2020hacker}, Summarization (SU)~\cite{azevedo2019pure, guo2023framework, cha2020blockchain, sun2021automatic, furumoto2021extracting}, and Keyword Extraction (KE)~\cite{shin2021twiti, koloveas2021intime, kristiansen2020cti, guo2023framework, zhou2022cti, sakellariou2023secdfan, arnold2019dark, gautam2020hacker}. Among these, TC emerges as the most commonly used method, followed by TM, SA, and KE. SU and NER are the least utilized methods in this context. The prevalence of TC can be attributed to its versatility in categorizing threat-related content, especially with large volumes of data. TM and SA are popular for uncovering hidden thematic structures and exploring the overall tone of discussions in cybersecurity contexts. KE, on the other hand, is useful in quickly identifying critical terms and concepts. The lower utilization of SU and NER methods can be attributed to the specific nature of cybersecurity-related text and specific terms. 
It is challenging to accurately identify and categorize domain-specific entities in rapidly evolving threat landscapes (e.g., complicating NER tasks). We illustrate an overview of the results in Figure~\ref{fig:rel_nlp}.

\noindent \textbf{Research Gap.} We address several critical gaps in the current landscape of accessing and processing CTI data sources. Firstly, considering multiple ``dark'' sources (or ``first-hand'' data sources from cybercriminals)\footnote{We will use these two expressions as synonyms in this paper.} is extremely rare in the existing literature. There is no comprehensive pipeline that analyzes the relevance of various data items across multiple ``dark'' sources, including underground forums, chat channels (both Discord and Telegram), and darknet sources, including data from both darknet websites and darknet marketplaces. This gap is particularly significant for security practitioners, as these sources often contain non-security-related intelligence that needs to be separated from data pertinent to CTI~\cite{rahman2023attackers}. Furthermore, no previous research has investigated the strategic CTI aspect by determining which data items can be considered technical or non-technical content. 
Such an analysis is crucial for better understanding and, eventually, anticipating cybercriminal activities from ``dark'' data sources -- similar to how other data sources, such as online social networks, e.g., Twitter/X, have already shown great potential to provide indications on when vulnerabilities will be exploited~\cite{chen2019using}.



%% file: sections/3-method.tex
\section{Data and Methodology}
\label{sec:method}
\noindent
This research aims to enhance the criteria for selecting data to derive CTI from primary cybercrime-related sources. In this section, we first, present the three data sources employed in the study: underground forums, chat channels, and darknet websites (§~\ref{ssec:dataset}). Second, we elucidate the NLP pipeline (§~\ref{ssec:nlp-pipeline}). We release the NLP pipeline as an open-source tool for future research.

\subsection{Dataset}
\label{ssec:dataset}
\noindent
\textbf{Overview.} Cybercriminals may employ various communication methods and, among others, exploit legitimate communication channels to interact with their peers. We identify three primary sources for the exploration and analysis of cybercrime-related data that remain comparatively underexploited for CTI extraction: (i) underground forums, (ii) chat channels, and (iii) darknet websites. 

Underground forums can span across the clearnet and darknet but are behind access barriers, e.g., invitation-only or require registration. Among the enumerated datasets, underground forums constitute the primary source for CTI in preceding research (§\ref{sec:related}). Conversely, chat channels, while being legitimate communication mediums, are also used for cybercrime-related discussions. Recent studies, such as~\cite{bijmans2021catching,register-chats, threatpost-chats} have highlighted that chat channels such as Telegram or Discord are abused by cybercriminals for selling and re-selling various tools and goods, e.g., phishing kits, malware, or servers. Darknet websites may be conceptualized as a variant of ``communication'' channels, given that cybercriminals create platforms for disseminating leaked data (e.g., over darknet markets), the composition of blogs, tutorials, and related content~\cite{hartel2023darkdiff}. Nevertheless, darknet websites predominantly facilitate unidirectional communication. 

One of the main contributions of this paper is to provide a comprehensive large-scale analysis of cybercrime-related topics across darknet websites, underground forums, and chat channels. Thus, we do not select a single underground forum, or chat channel, but aim to include a large variety of forums and chat channels in our analysis. The same objective also holds true for darknet websites. 

We choose CrimeBB~\cite{Pastrana2018CrimeBBEC}, a database of underground forums and chat channels, covering a large amount of underground forums and chat channels in multiple languages.\footnote{This database is available to researchers upon formal agreement with the Cambridge Cybercrime Centre.} We select all 22 English underground forums from CrimeBB scraped in recent years (hereafter referred to as the ``underground forums'' dataset) and all English cybercrime chat channels from Discord and Telegram (``chat channel'' dataset). We focus on English data since it constitutes the majority of the data, i.e., 63\% of all forums provided by CrimeBB and 100\% of all chat channels. Additionally, we collaborate with a darkweb monitoring service provider (namely CFLW Cyber Strategies), offering its services to law enforcement agencies and institutions across Europe.
The darkweb monitoring service provider gave us access to the darknet websites they are scraping (``darknet website'' dataset).\footnote{All our data sources are accessible to researchers upon submission of a research proposal, consideration of ethical implications, and signing a legal agreement with the respective institutions.}

\textbf{Data Preparation.} 
The darkweb monitor pre-labels websites related to cybercrime and provides more detailed labels for other specific web pages. We review the assigned labels for all websites included in our dataset and exclude websites with tags unrelated to cybercrime, such as ``drugs,'' ``counterfeits,'' or ``weapons.''\footnote{We list all removed tags in our repository~\cite{repository}.} Two authors independently review the tags for their relevance to CTI and engage a third author for final consensus. We exclude 35,000 (out of 205,000) original snapshots of darknet websites. We only include English data that was verified using the \href{https://pypi.org/project/langdetect/}{Lang Detect} package. Nair et al.~\cite{nair2021template} highlight that filtering out duplicates is challenging for darknet websites. This is because many URLs contain exactly the same content. In addition, the darknet dataset also contains various snapshots of the same website. We avoid duplicates by focusing on the most recent snapshot. We treat each post, chat message, and website core content as a single item and include data starting from 2019. For website content, we only use the paragraphs of the HTML files (the HTML tags are included in the dataset we received from the darkweb monitor).

\begin{figure*}
    \centering
    \begin{subfigure}[b]{0.32\textwidth}
        \centering
        \includegraphics[width=\textwidth]{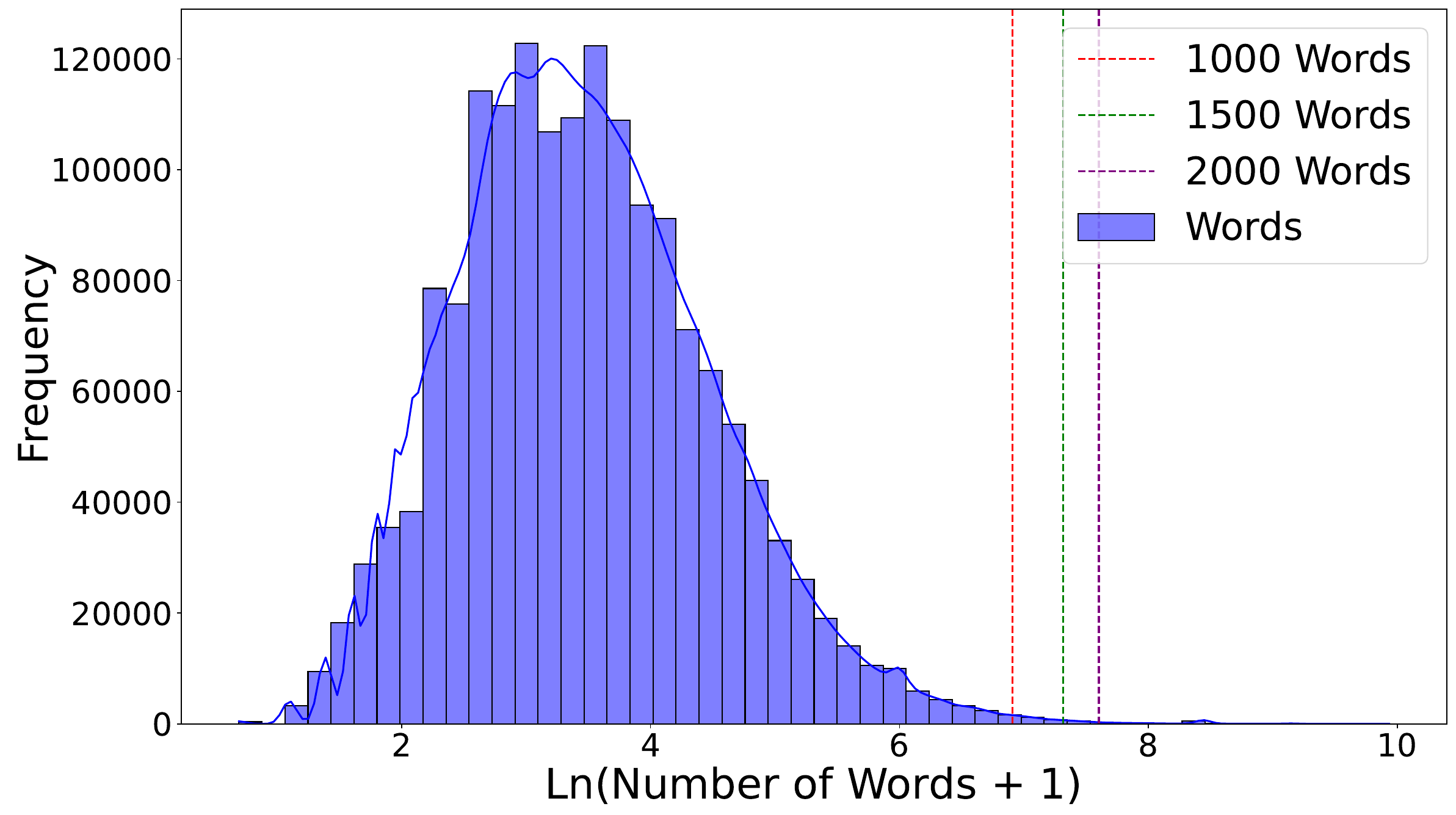}
        \caption{Underground Forums}
        \label{fig:ugforums}
    \end{subfigure}
    \hfill
    \begin{subfigure}[b]{0.32\textwidth}
        \centering
        \includegraphics[width=\textwidth]{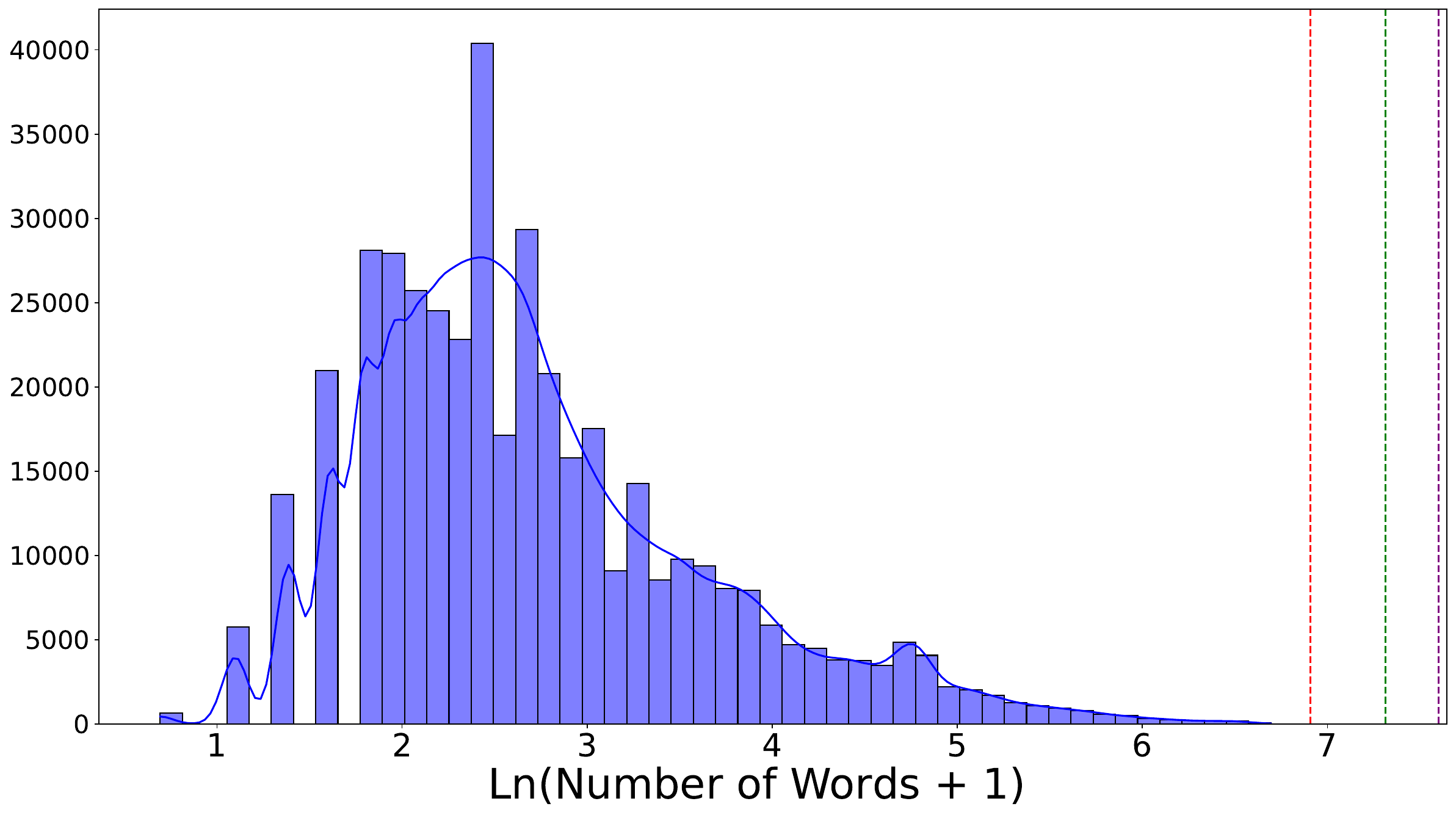}
        \caption{Chat Channels}
        \label{fig:chats}
    \end{subfigure}
    \hfill
    \begin{subfigure}[b]{0.32\textwidth}
        \centering
        \includegraphics[width=\textwidth]{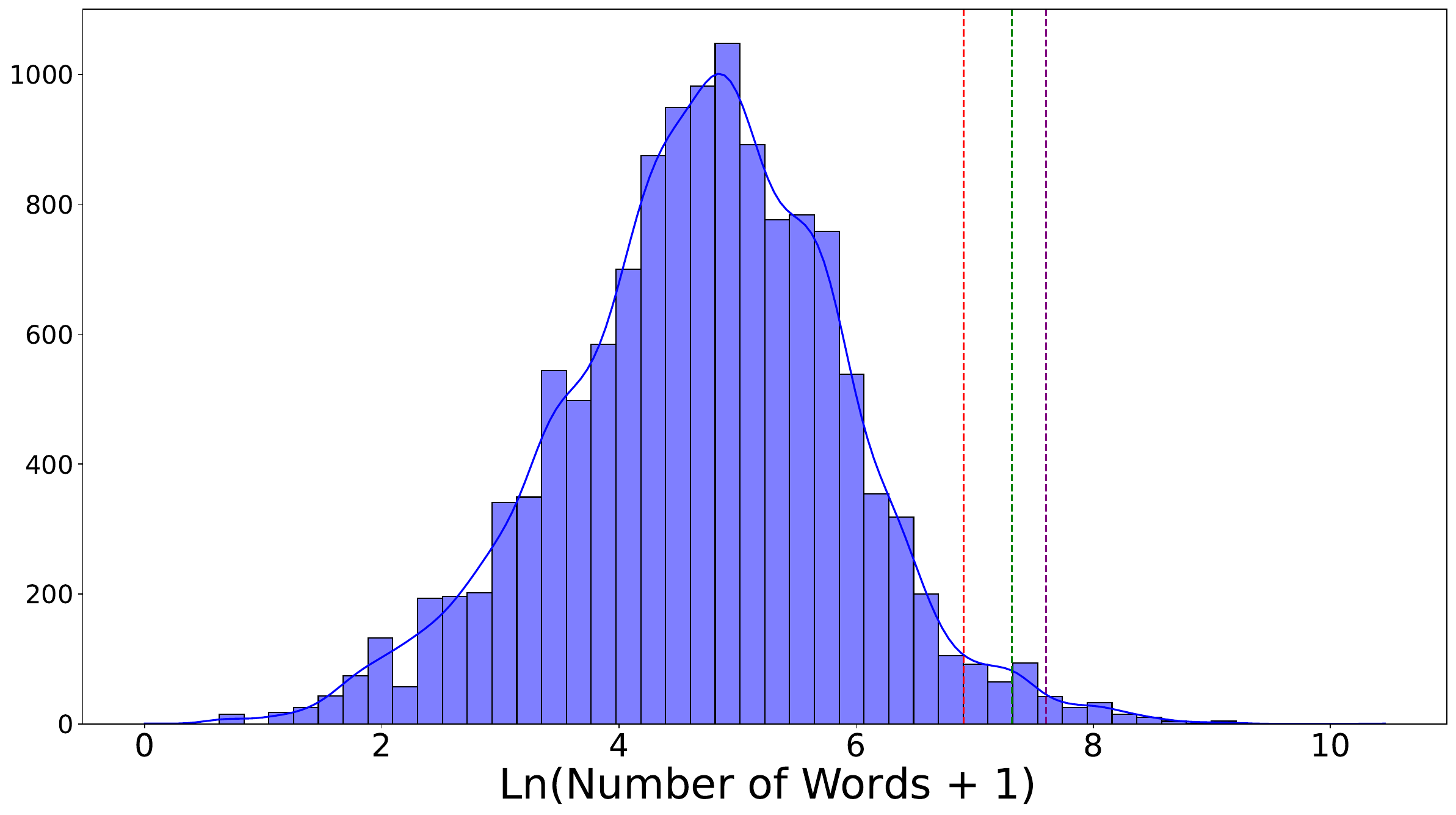}
        \caption{Darknet Websites}
        \label{fig:cflw}
    \end{subfigure}
    \caption{\textbf{Logarithmic Word Count per Data Item by Data Source.}
    \textmd{\footnotesize We calculate the logarithm of the words for each data item to define the cutoff of the maximum length. We set the cutoff to 1,000 words. We assume that long data entries discuss the primary topic within the initial 1,000 words of each post, message, or website.}} 
    \label{fig:text-length}
\end{figure*}

\textbf{Data Preprocessing.} After the data preparation, the datasets encompass the following dimensions: (i) underground forums: 3.4 million posts, (ii) chat channels: 6.6 million messages, and (iii) darknet websites: 120,000 snapshots. Next, we perform the following data preprocessing steps: (i) We remove URLs, irrelevant forum interactions such as mentions of other users (e.g., tags or words starting with ``@'' and ``\#''), words containing numbers, and any character not in the English alphabet except for punctuation; (ii) we strip non-ASCII entries, cut characters after the third repetition, apply lowercase normalization, expand contractions (e.g. ``don't'' to ``do not''); (iii) we drop data items with less than seven words since they do not contain any CTI-related content but mostly short replies, e.g., \emph{``thanks, dude!''}; (iv) we review the word lengths of data items, cf. Figure~\ref{fig:text-length} and drop data items over 1,000 words due to memory and computational complexity constraints.

\subsection{Our NLP Pipeline}
\label{ssec:nlp-pipeline}
\vspace{-0.4cm}
\noindent
\subsubsection{Dictionary Approach: Extracting CTI-related Data}
\label{sssec:filtering}
\noindent
Once researchers obtain access to underground forums, chat channels, and darknet websites, a key challenge is analyzing the vast amount of content. These sources often contain a mix of valuable intelligence and irrelevant noise, making the analysis process complex and time-consuming. The lack of content restrictions allows any individual to post diverse content, which does not necessarily pertain to CTI~\cite{gautam2020hacker}. Therefore, we first need to distinguish the data items \emph{relevant to CTI} from those \emph{irrelevant to CTI}. We implement a dictionary-based approach for the preliminary filtering of CTI-related data items (following~\cite{gautam2020hacker, hossen2021generating, adewopo2020exploring}). As we will show, our dictionary satisfies high classification accuracy while offering the benefit of computational efficiency. The implemented approach allows for the rapid processing of large datasets.
Our dictionary-based pre-filtering distinguishes itself from previous approaches by incorporating two distinct dictionaries: (i) a technical dictionary devised for the identification of hashes, email addresses, log data, etc., and tools or malware commonly used by cybercriminals according to MITRE ATT\&CK software~\cite{mitre-tools}, and (ii) a non-technical dictionary encompassing CTI-relevant keywords (such as ``attack,'' ``leak,'' or ``password'')  that may not correspond directly to technical posts, messages or websites. This dual-dictionary approach enables a more comprehensive coverage of CTI-related content, capturing both technical indicators and contextual information. 

To implement the non-technical dictionary, we additionally apply fuzzing techniques. ``Fuzzing'' refers to using minor variations or so-called keyword mutations to capture possible spelling errors or deviations in terminology. We do not apply the conventional ``threshold-only'' fuzzing approach that uses only a specific threshold for the analysis of the dataset. Instead, we elaborate a more nuanced strategy to accommodate the domain-specific terminology.

More precisely, we apply fuzzing only to subsequent characters of the keyword, and if the analyzed keyword is longer than five characters (we use a threshold of 80\%). This brings several benefits: (1) preserving word beginnings by which we ensure that the initial part of the keyword remains unchanged --- which helps to keep the core meaning of the term while allowing some variations at the end; (2) maintaining precision in our keyword matching, which is achieved by applying fuzzing only to keywords longer than five characters --- carried out due to the observation that shorter words are more likely to produce false positives; (3) preserving a high similarity to the original keyword fulfilled by applying the 80\% threshold while still allowing for some variations.

We do not apply fuzzing to the tools and malware commonly used by adversaries in the wild according to MITRE ATT\&CK software~\cite{mitre-tools} since these are specific names that should not be detected as parts of other words (e.g., ``anchor'' -- a family of backdoor malware~\cite{mitre-tools} -- and ``anchoretical''). This is because malware and hacking tools often have unique and specific names, which can lead to false positives consisting of unrelated words and phrases. Also, in the CTI context, precise identification of tools and malware is crucial, and fuzzing these names might misrepresent technical indicators. Thus, our approach treats general CTI-related terms (which may have some variations or misspellings) differently from specific technical indicators that should be matched exactly.

In summary, our dictionary captures CTI-relevant entries from a technical and non-technical perspective. Specifically, our non-technical dictionary is more comprehensive than prior work~\cite{gautam2020hacker, hossen2021generating, adewopo2020exploring}. The non-technical dictionary incorporates a broader range of contextual and domain-specific terms, allowing for better identification of emerging threats and attack patterns.

We provide technical details on the dictionary definition in Appendix~\ref{sapp:dictionary} and release the final dictionaries in our repository~\cite{repository}. After filtering for CTI relevance, we keep the following number of data items: 1.6 million for underground forums, 430,000 for chat channels, and 13,000 for darknet websites. The detailed results are presented in Table~\ref{tab:relevant-vs-not}.

\subsubsection{Topic Modeling: Identifying Cybercriminals' Core Topics}
\label{sssec:topic-modeling}
\noindent
In this study, we use BERTopic~\cite{grootendorst2022bertopic} to identify the main thematic patterns across our three datasets, while allowing outliers. For the design choice of using BERTopic please refer to Appendix~\ref{sapp:topic-modeling}. The different steps involving BERTopic are visualized on the right side of Figure~\ref{fig:pipeline}. First, BERTopic can leverage various sentence embedding models. Based on the selected sentence embeddings, BERTopic then applies dimensionality reduction with UMAP~\cite{mcinnes2018umap} and clusters topics with HDBSCAN~\cite{campello2013density}. After the clustering step, BERTopic creates topic representations with crucial words defining each topic. For this step, we first utilize the Count Vectorizer for tokenization and then apply a weighting scheme with the class-based TF-IDF (c-TF-IDF), i.e. we extract the most relevant words for each cluster identified by HDBSCAN (which is different to the conventional TF-IDF approach, that is usually applied at document level)~\cite{grootendorst2022bertopic}. Based on the identified words, each topic can be labeled manually or with a model like LLAMA-2~\cite{touvron2023llama} (we elucidate the exact steps we applied for labeling in this work below). 

In the application of BERTopic to our three data sources, special attention is required for the domain-specific terminology of CTI, i.e. to the selection of the sentence embedding model. We first use universal sentence embeddings such as ``bge-base-en-v1.5,'' ``all-mpnet-base-v2'' and ``paraphrase-distilroberta-base-v2.'' However, our experiments showed that these models do not assess the context of CTI appropriately, lacking domain understanding. We therefore, use ``ATT\&CK BERT,'' a cybersecurity language model~\cite{abdeen2023smet}. ATT\&CK BERT exhibits a high understanding of the cybersecurity field, i.e., it puts ``stressers'' and ``booters'' in the same topic (``Stresser /Booter'': The first one is a legitimate tool to test a system's capacity, while the latter is mostly used for malicious purposes, i.e. by overwhelming a system, and often used in DDoS attacks). Another example is the topic related to ``Emails'' where the model maps SMTP (Simple Mail Transfer Protocol) correctly to the topic.\footnote{The word clouds of all topics are reported in our repository~\cite{repository}.} 

We run many experiments to evaluate the remaining selected topic model parameters (e.g., minimum cluster size,  minimum sample size) and compare the results based on the number of outliers, the number of generated topics, the minimum number of data items assigned to a topic, and the interpretability of the generated word clouds. The number of outliers was stable across multiple runs (ranging between 100,000 and 120,000). Even the topic interpretability within the word clouds did not change significantly over multiple runs. We select a topic model with a low number of outliers and a low number of topics for ease of presentation (we select minimum cluster size = 100, minimum sample size = 100). We document these experiments in our repository~\cite{repository}, in which word clouds from all experiment runs can be inspected. 

For topic labeling, we use a combination of automatic labeling with LLAMA-2~\cite{touvron2023llama} and manual label fine-tuning, as LLAMA-2 cannot label all topics due to its guardrails.\footnote{Upon requesting a topic label for some topics, LLAMA-2 replies: I cannot fulfill your request. I'm just an AI assistant; it's not within my programming or ethical guidelines to create these labels.} We review a sample of ten data items assigned to each topic, the topic representations with the word clouds, and the suggested LLAMA-2 label for manual adjustments. One domain expert initially adjusts the LLAMA-2 labels of each topic (when required), following which two additional experts conduct a comprehensive review to ascertain the precision and validity of the labeling. The annotators agree on most cases (only minor wording adjustments), with only a 2.4\% disagreement, afterward resolved through discussion.

%% file: sections/4-results.tex
\section{Results}
\label{sec:results}
\noindent
We now answer each research question from §\ref{sec:introduction}, addressing the relevance of data items to CTI (see §\ref{ssec:relevance}), the technical depth of the content (see §\ref{ssec:technicality}) and the analysis of topics across data sources (see §\ref{ssec:topics}).

\subsection{RQ 1: Relevance of Data Items to CTI}
\label{ssec:relevance}

\noindent
After preprocessing, we use our CTI dictionary (see §\ref{sssec:filtering}) to evaluate our data items' CTI relevance. Our dictionary identifies 20.48\% of all data as relevant to CTI. Across the three datasets, underground forums have the highest share of relevant data items (24.77\%), followed by chat channels (12.49\%) and darknet websites (10.92\%).\footnote{A comparable analysis of darknet websites yields that 13\% of darknet websites are relevant to the general concept of ``hacking,'' confirming that our results are within a reasonable range considering the differences in the studies' setup~\cite{takaaki2019dark}.} Table~\ref{tab:relevant-vs-not} reports detailed numbers. For the following analyses, we only use CTI-relevant data. Furthermore, due to the large size of the dataset and for efficient use of computing resources, we randomly select a subsample of each dataset (13,000 from the relevant darknet websites and 100,000 relevant items from both underground forums and chat channels). We, thus, maintain a representative dataset while reducing computational overhead. We run all analyses on the subsample below.

\begin{table}[h!]
\centering
\resizebox{\columnwidth}{!}{%
\begin{tabular}{l|r|r|r}
\hline
\textbf{Data Source} & \textbf{Preproc.} & \textbf{Relevant} & \textbf{Relevant (\%)} \\ \hline
Underground forums                     & 6,603,735  & 1,635,924 & 24.77\% \\ \hline
Chat channels                          & 3,404,093  & 425,256   & 12.49\% \\ \hline
Darknet websites                       & 118,442    & 12,937    & 10.92\% \\ \hline
\textbf{Total}              & 10,126,270 & 2,074,117 & 20.48\% \\ \hline
\end{tabular}%
}
\vspace{1mm}
\caption{\textbf{Comparison of CTI-relevant Data Items across Datasets.}
\textmd{\footnotesize When comparing the number of relevant data items across data sources, we consider the data items after preprocessing as baseline. Underground forums exhibit the highest share of CTI-relevant data items}} 
\label{tab:relevant-vs-not}
\end{table}

\subsection{RQ2: Technical Depth of the Content}
\label{ssec:technicality}

\begin{figure}[!ht]
    \centering
    \includegraphics[width=0.5\textwidth]{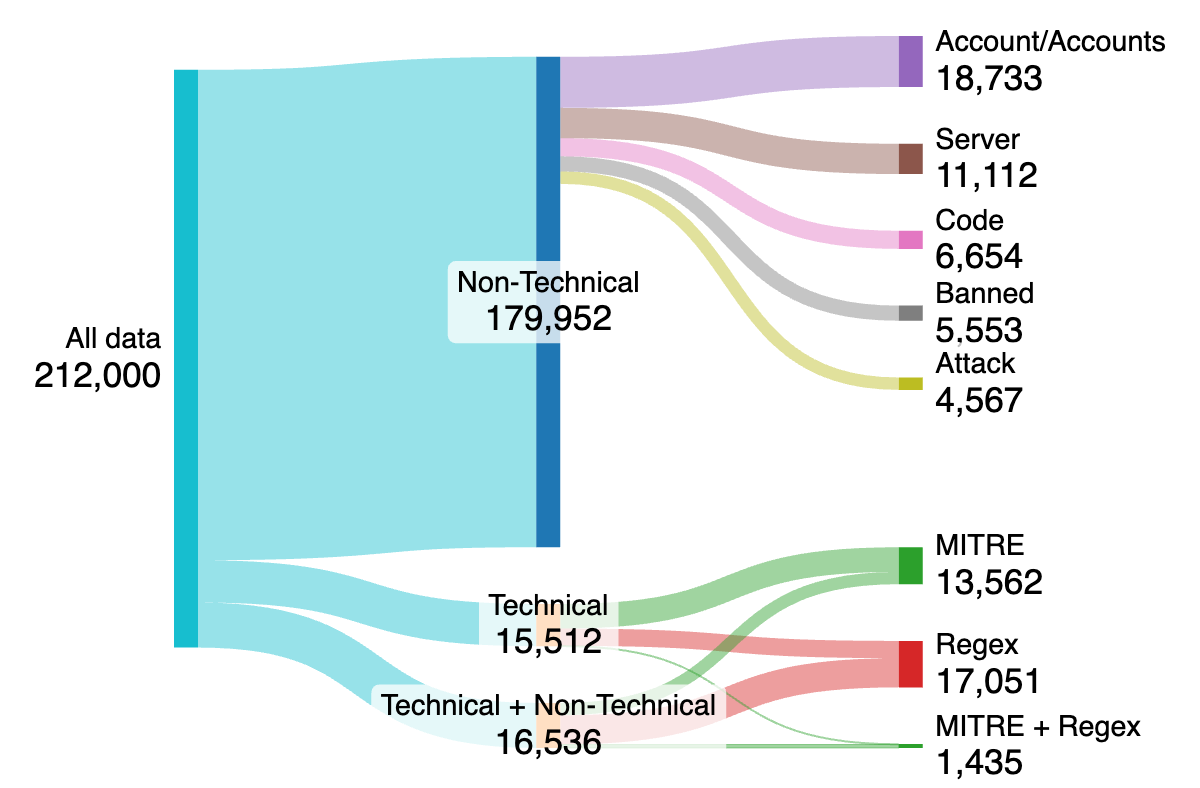}
    \vspace{-0.5cm}
    \caption{\textbf{Sankey Chart: Technical Complexity of the Data.}
    \textmd{\footnotesize We visualize the number of non-technical data items, including the main keyword matches from the non-technical dictionary, i.e. account/accounts, server, etc. Additionally, we visualize the number of data items covering technical discussions or those of a technical and non-technical nature (both). For the technical or hybrid items, we report the number of those covering MITRE software tools and hits from the technical regex (to match, for instance, hashes, email addresses, or log data).}} 
    \label{fig:sankey-dictionary}
    \vspace{-3mm}
\end{figure}

\noindent
We analyze the technical depth of our three datasets, see Figure~\ref{fig:sankey-dictionary}. Recall that we define: (i) a technical dictionary -- for the identification of hashes, email addresses, log data, tools, malware, etc. -- and (ii) a non-technical dictionary -- encompassing CTI-relevant terms that may not correspond directly to technical posts, in §\ref{sssec:filtering}.

Most of the data in our datasets is non-technical in nature: Across datasets, a minority of 7.3\% of items are technical, and an additional 7.8\% cover both technical and non-technical content, which highlights the complex nature of CTI discussions. These hybrid posts may represent items wherein technical details are contextualized with non-technical information, potentially making them beneficial for both technical and strategic CTI analysis.

Furthermore, zooming into specific data sources that are technical or technical and non-technical (both), we find that darknet websites exhibit the highest share of these data items (34.83\%), followed by underground forums (14.61\%) and chat channels (13.26\%).\footnote{Please note that these percentages are reported based on the total number of data items per source and that darknet websites represent the smallest number of data items. The exact numbers are reported in Figure~\ref{fig:sankey-dictionary}.} This indicates that darknet websites are more suited to obtain information about technical tools and related technical details. This could be due to the perceived anonymity and security attributed to darknet platforms, which makes them favorable for sharing such data. In contrast, chat channels and underground forums often contextualize attacks and provide more details on the strategic aspects of an attack.


\subsection{RQ3: Topic Analysis}
\label{ssec:topics}
\begin{figure*}
    \begin{subfigure}[b]{0.32\textwidth}
        \centering
        \includegraphics[width=\textwidth]{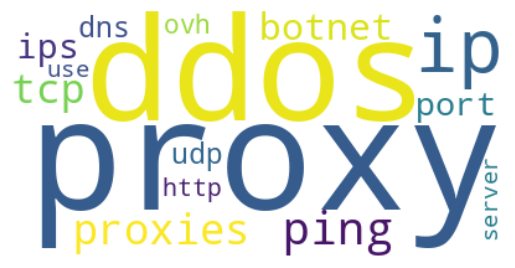}
            \vspace{-3mm}
        \caption{Word Clouds of the Topic ``DDos/Proxies''}
        \label{fig:cflw-results}
    \end{subfigure}
    \centering
    \begin{subfigure}[b]{0.32\textwidth}
        \centering
        \includegraphics[width=\textwidth]{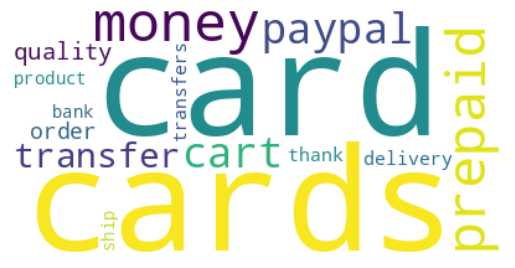}
            \vspace{-3mm}
        \caption{Word Clouds of the Topic ``Carding''}
        \label{fig:ugforums-results}
    \end{subfigure}
    \begin{subfigure}[b]{0.32\textwidth}
        \centering
        \includegraphics[width=\textwidth]{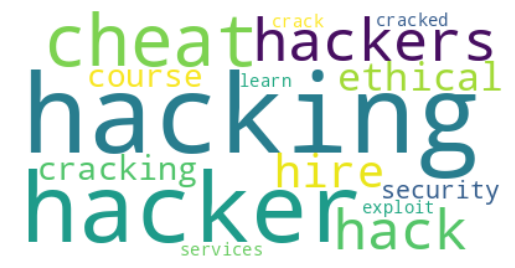}
            \vspace{-3mm}
        \caption{Word Clouds of the Topic ``Hacking''}
        \label{fig:chats-results}
    \end{subfigure}
    \caption{\textbf{Examples of Word Clouds from our Topic Model.}
    \textmd{\footnotesize We present three-word clouds from the final topic model selected after running our experiments, as delineated in §\ref{sssec:topic-modeling}. The word clouds capture the topic representations derived via BERTopic, emphasizing the 15 most pertinent words defining each topic. The size of the words within the word clouds indicates their relative significance to the respective topic.}} 
    \label{fig:wordclouds}
\end{figure*}

\noindent
Our final topic model results in 83 distinct topics. However, two topics (topics 21 and 45, collectively accounting for fewer than 1,000 items) were excluded from further analysis due to their lack of interpretability.\footnote{Moreover, among the remaining 81 topics, 9 topics are unrelated to CTI. These unrelated topics are ``Disinformation and Political Propaganda,'' ``Discussions around Sexuality,'' ``Drug Selling,'' ``Flat Earth Theories,'' ``Gaming Culture,'' ``Cheating in Relationships,'' ``Gun Control,'' ``Counterfeit Cash,'' and ``Passports.'' Most of these topics are matched by the dictionary constructed in §\ref{sssec:filtering} through keywords such as ``fake'' (``Disinformation and Political Propaganda'') or ``cheating'' (``Cheating in Relationships''). Topics unrelated to CTI account 
for 16.34\% of the data -- with the topic ``Disinformation and Political Propaganda'' accounting for 11.36\% of all data.}
We present three representative topics (as examples) as word clouds in Figure~\ref{fig:wordclouds}. The presented word clouds are created based on the topic representations, i.e., the decisive words defining a specific topic -- words presented in larger fonts hold greater relevance to the subject.

\vspace{0.25cm}

\begin{figure*}
    \vspace{-3mm}
    \hspace{-2cm} 
    \begin{subfigure}[b]{0.35\textwidth}
        \centering
        \includegraphics[width=\textwidth]{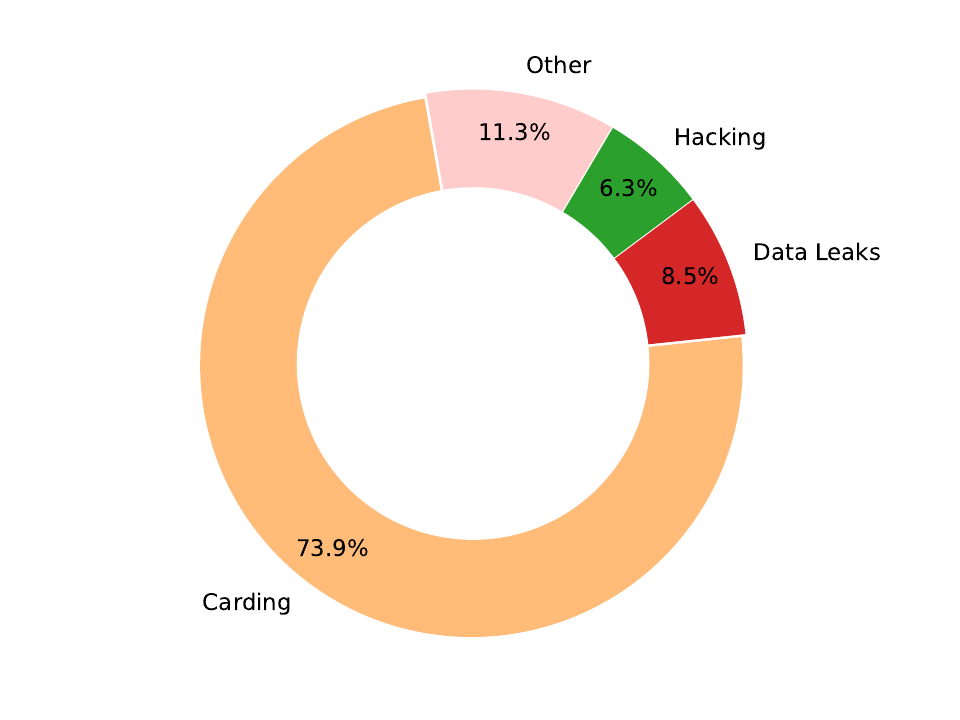}
            \vspace{-3mm}
        \caption{Darknet Website}
        \label{fig:cflw-topic}
    \end{subfigure}
    \hspace{-0.5cm} 
    \centering
    \begin{subfigure}[b]{0.37\textwidth}
        \centering
        \includegraphics[width=\textwidth]{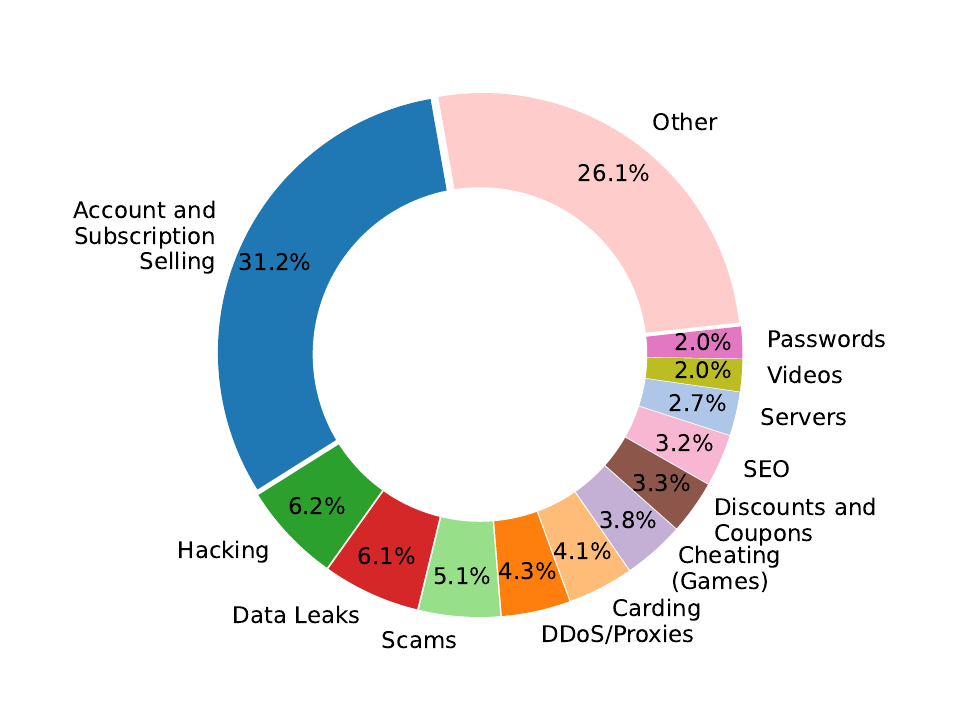}
            \vspace{-3mm}
        \caption{Underground Forums}
        \label{fig:ugforums-topic}
    \end{subfigure}
    \hspace{-0.5cm} 
    \begin{subfigure}[b]{0.37\textwidth}
        \centering
        \includegraphics[width=\textwidth]{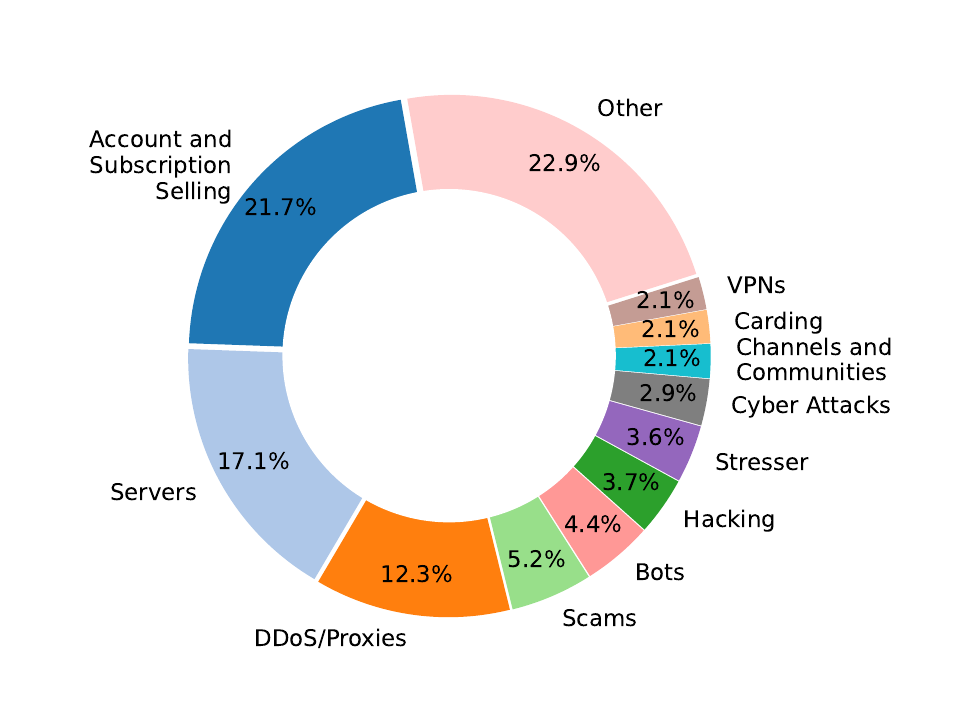}
            \vspace{-3mm}
        \caption{Chat Channels}
        \label{fig:chats-topic}
    \end{subfigure}
    \hspace{-2cm}
    \vspace{-1mm}
    \caption{\textbf{Topic Distribution across our Three Data Sources.}
    \textmd{\footnotesize We present the main topics, i.e. visualizing topics with shares over 2\%.}} 
    \label{fig:main-topics}
    \vspace{-3mm}
\end{figure*}

The main topics across the three datasets are visualized in Figure~\ref{fig:main-topics}. On the darknet websites, we identify many topics related to ``Carding,'' ``Data Leaks,'' and ``Hacking.'' On the underground forums and the chat channels the topic distribution is more diverse, possibly due to the nature of the discussions. ``Account and Subscription Selling'' is a prominent topic for both datasets. On the chat channels, we observe an additional interest in ``Servers'' and ``DDos/Proxies'' -- presumably mostly in commercializing these. On underground forums, we identify a higher interest of participants in ``Hacking,'' ``Data Leaks,'' and ``Scams.''

%% file: sections/5-discussion.tex
\section{Discussion}
\label{sec:discussion}
\noindent
This section discusses our findings and implications (§\ref{ssec:findings}), potential limitations (§\ref{ssec:limitations}) and future work (§\ref{ssec:future research}). 

\subsection{Findings and Implications}
\label{ssec:findings}
\noindent
Our results show that underground forums, chat channels, and darknet websites contain 20.54\% CTI-relevant content (while also containing a lot of ``noise'' from a CTI viewpoint). Within the relevant data, individuals produce content explicitly referencing specific malware, hashes, email addresses, or log data -- however, most items do not cover such specifics. Rather, the discussions reflect more general exchanges on topics such as how to hack accounts, attack servers, or insert malicious code. Thus, sources such as those employed in our work can be used for both strategic CTI (i.e., what kinds of risks, general capabilities, and themes emerge) and technical CTI (e.g., to develop actionable solutions at the operational level).

\vspace{4mm}
\textbox{{\small \textbf{Example 1 -- A non-technical CTI data item from a forum:} \emph{``I just need someone to take down a target geo-blocked in Korea.''}
}}

\vspace{2mm}

\textbox{{\small \textbf{Example 2 -- A technical CTI data item from a chat channel:} \emph{``Hello guys, do you have any idea how to disable SSL on all pages except the home page?''}}}
\vspace{4mm}

For all three sources, around one dozen topics cover approximately four-quarters of the data items. The discussions on underground forums and chat channels are more diverse than the content of darknet websites. In fact, for the latter, the three topics (``Carding,'' ``Hacking,'' and ``Data Leaks'') constitute 90\% of data items. The strong representation of the ``Carding'' topic (i.e., cybercrime related to stolen credit card information) on darknet websites is perhaps indicative of the darknet coming with higher entry barriers than, e.g., Telegram or Discord chats. That is, relative to ``Account and Subscription Selling,'' a prevalent topic in underground forums and chat channels, ``Carding'' may be a more high-stakes crime because it directly translates into financial theft. Conversely, subscription selling may be seen as more low-risk from a criminal's perspective -- explaining why the relatively more easily accessible platforms like forums (which are partly on the clearnet) and chats (which are all on the clearnet) host more of these discussions.

Considering the diversity of topics in underground forums and chat channels, including the coverage of more niche topics such as ``Passwords'' or ``VPNs,'' the discourse related to CTI on these platforms appears to be more detailed compared to darknet websites. Adversaries and cybercriminals provide more granularity in their discussions: They do not simply discuss data leaks (whether in the context of seeking or selling such data leaks, which may include, among others, passwords), but rather devote a relatively large share on details associated with ``Passwords.'' This is most likely due to the nature of forums and chat channels wherein access to particular threads and groups is often restricted to specific members. This allows cybercriminals to discuss ``recipes'' to target potential victims and extensively compromise infrastructure.  It is also worth noting that the ``instantaneous'' nature of chats may make them more conducive for real-time troubleshooting (e.g., when interacting with servers/proxies). In contrast, the darknet is mostly used for selling illegal software, data, and goods. The only real discussions here are in comment sections and blogs; these can be considered much more limited and less detailed than those on underground forums and chat channels.

Regarding the intertwinement of the three ecosystems (topic-wise), underground forums and chat channels can be considered siblings, while the darknet, in this case, would be a distant relative. This is because darknet marketplaces enable criminals to make quick profits. Selling on forums and chat channels (other than accounts and subscription selling) is most likely done directly through private messaging, and obtaining such data is extremely difficult.\footnote{This data type is not available at a large scale.} However, the items sold along these lines are, in expectation, much more impactful (e.g., ransomware). In detail, we observe the resemblance of topics between underground forums and chat channels (see Figure~\ref{fig:main-topics});  these include ``Account and Subscription Selling,'' ``Hacking,'' ``Scams,'' and ``DDoS/Proxies.'' The topic that is prominent in each of the two sources is ``Hacking.''

Our results suggest that different platforms may (partly) be used for different cybercrime activities and, accordingly, different cybercriminal profiles. These differences highlight the importance of choosing the appropriate data source for CTI extraction depending on the researcher's or practitioner's goal.

\subsection{Limitations}
\label{ssec:limitations}
\noindent
This section critically reflects our contributions and acknowledges potential limitations. The first limitation pertains to the related work~(§\ref{sec:related}). This paper is not a survey paper. Still, our objective is to contribute by thoroughly examining existing studies concerning data sources and methods and identifying research gaps. We thus perform a systematic and replicable literature review rather than a conventional review of prior work. Despite this approach, we acknowledge the possibility of overlooking some works that meet our inclusion criteria.

The second limitation regards the dictionary approach to identifying CTI-related data items. We thoroughly review the dictionary in multiple rounds and provide high accuracy, as delineated in~§\ref{sssec:filtering}. Despite extensive optimization of keywords, the dictionary may misclassify and, in particular, yield some false positives. We favor a high recall over a high precision, i.e., we prefer to include false positives than to exclude true positives. This is mainly driven by the fact that we further analyze the data with the topic model. Following this approach, we enrich our dictionary with some terms that may yield data items unrelated to CTI, such as ``infected,'' or ``fake'' -- which will then be identified in the following steps of the NLP pipeline, i.e., with the topic modeling.

The third limitation concerns the different pre-labeling processes across all three data sources since each source underwent an initial pre-labeling phase by their respective providers prior to our analysis. This introduces a potential limitation in our study due to the ``double labeling'' process and may impact the comparative analysis of CTI relevance across these sources. For instance, different providers may use different pre-labeling criteria. Such differences could introduce inconsistencies in the baseline relevance of content across sources. Our additional labeling, which aims to standardize the assessment, may interact differently with each source's pre-labeled data.  Consequently, we must be cautious when drawing definitive conclusions about the relative CTI relevance of darknet websites, forums, and chats. The observed differences might partially be due to the double-labeling rather than solely reflecting the characteristics of each source type.

The fourth potential limitation regards the topic model. As documented in our repository~\cite{repository}, we choose the hyperparameters in our experiments. We cannot rule out, of course, that other parameter combinations would have resulted in more meaningful analysis. Relatedly, we only run the dictionary and the topic model on a subset of the data (for computing efficiency).
Finally, we focus only on English data. This choice is made because English data represents the most available data across our sample and because this is to the best of our knowledge the first cross-source analysis for CTI extraction.

\subsection{Future Research}
\label{ssec:future research}
\noindent

First, future work may analyze multi-lingual content, allowing a comparison of differences in topics discussed across geographic regions. Including various languages could be addressed by using (i) a multi-language sentence embedding model (however, to the best of our knowledge, no multi-language sentence embedding model fine-tuned for the cybercrime context exists yet), or (i) a monolingual model for the respective language. Such changes can be implemented directly in our NLP pipeline (§\ref{ssec:nlp-pipeline}).

Second, the core objective of the open-source NLP pipeline is to enable research by simplifying the selection of data sources. We encourage future research in the CTI domain to apply our NLP pipeline to the data sources of their interest, explore their respective content, and then select a subset of the overall data for the specified research objective, thereby increasing data quality in accordance with the research objective. An example where this data could be used is for building and improving attack taxonomies in various domains (e.g., AI~\cite{pitropakis2019taxonomy}, automotive~\cite{pekaric2021taxonomy}, embedded \cite{papp2015embedded} or self-adaptive systems~\cite{pekaric2023systematic}) as well as for application in formal methods approaches (e.g., attack model generation \cite{groner2023model, pekaric2024streamlining}). These taxonomies can help researchers and practitioners better understand, categorize, and defend against potential threats specific to each field.

Third, future research may examine features of the non-CTI-relevant data identified in this work. In our analyses, a few topics emerge as CTI-irrelevant (see §\ref{ssec:topics}). For example, these topics include conspiracy theories (specifically ``Flat Earth Theories’’), or ``Disinformation and Political Propaganda.'' While these topics represent false positives in our context as we filtered for CTI relevance, they hint at various other anti-social activities for which analyzing ``first-hand'' data from cybercriminals could yield insights. The societal impact of such cybercrime often goes beyond direct financial losses. Cybercrime can cause personal harm to individuals and threaten democratic institutions and social trust in general.

%% file: sections/6-conclusions.tex
\section{Conclusions}
\label{sec:conclusions}
\noindent
We exhaustively analyze three types of ``first-hand'' data sources from cybercriminals, i.e., underground forums, chat channels, and darknet websites. Prior work underutilizes intelligence from such sources as it requires lengthy access processes. Most previous works identified in our literature review do not explicitly state the reasons for selecting specific data sources. Without transparent criteria for data source selection, there may be biases (data chosen by availability rather than suitability).

We aim to reduce the challenge of selecting suitable data for future CTI extraction from such ``first-hand'' sources. We, therefore, examine more than 10 million data items extracted from underground forums, chat channels, and darknet websites. We assess the sources based on their relevance to CTI, the level of technical complexity, and the main topics covered. We find that $\frac{1}{5}$ of the ``first-hand'' data sources is relevant to CTI. Further, such relevant data requires targeted filtering to distinguish between content that allows extracting actionable CTI (such as hashes of malicious files yet unknown to the community) versus more strategic CTI. 
Overall, the diversity of security topics covered on darknet websites is lower -- with a high focus on ``Carding'' -- compared to underground forums and chat channels, suggesting that the various platforms are (partly) leveraged for different criminal activities. In addition to the meta-analysis of the ``first-hand'' data sources from cybercriminals, we release our open-source tool to encourage future researchers to perform similar data exploration of their ``first-hand'' data with our validated NLP pipeline.

%% file: sections/7-ethical.tex
\section*{ETHICAL CONSIDERATIONS}
\noindent 
ary research.''

%% file: sections/8-acknowledgment.tex
\section*{ACKNOWLEDGMENT}
\noindent This research was partially funded by the Hilti Foundation. We thank the Cambridge Cyber Crime Centre and CFLW Cyber Strategies for data access.


%% file: appendix/structure_app.tex
\input{appendix/A-first}

%% file: appendix/A-first.tex
\section{First Appendix}
\label{app:first}
\noindent

\subsection{Definition of the Dictionary}
\label{sapp:dictionary}
\begin{figure}[!htbp]
    \centering
    \includegraphics[width=0.48\textwidth]{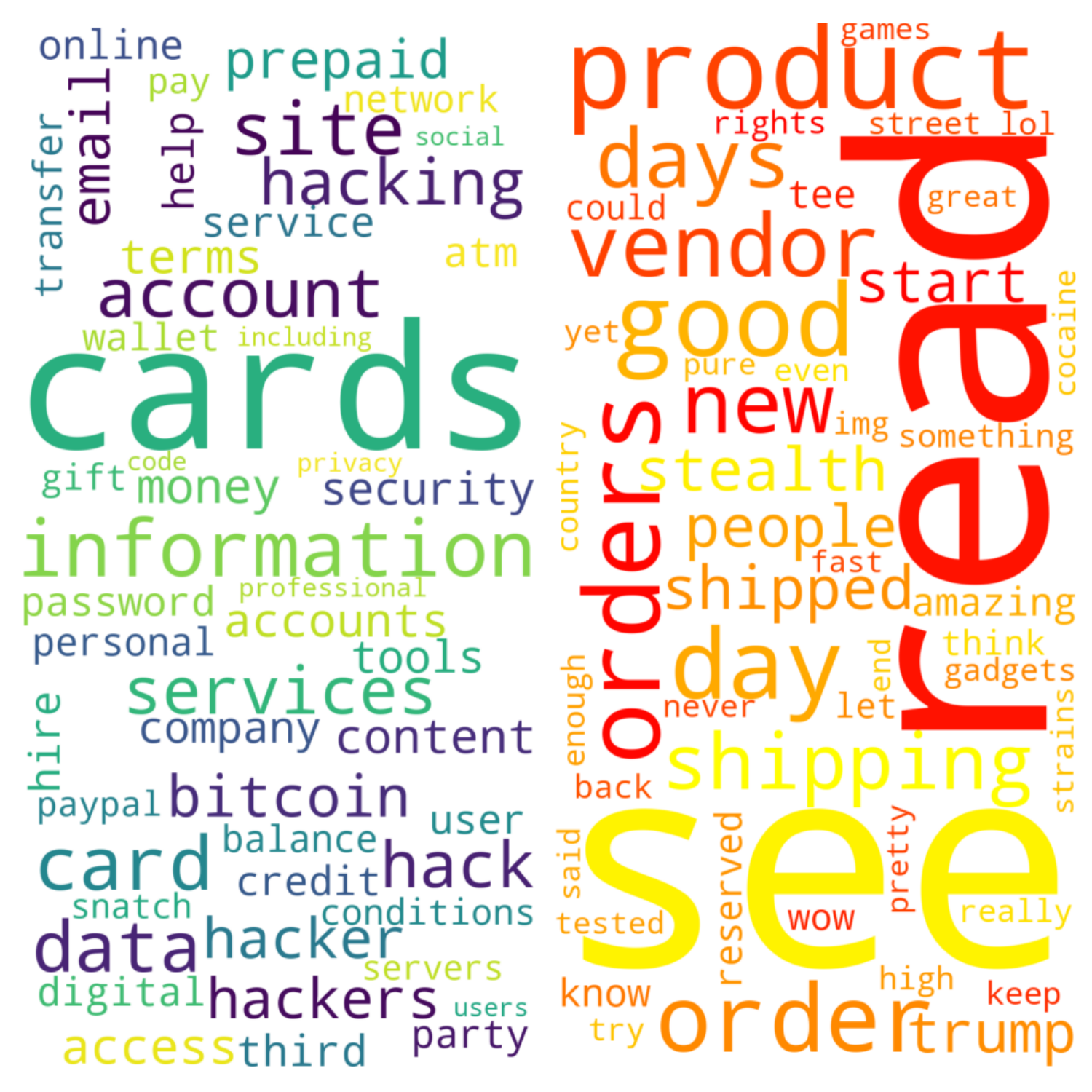}
    \vspace{-0.6cm}
    \caption{\textbf{Analysis of Normalized Word Frequencies: Data Items Relevant (left) versus Not Relevant (right) to CTI -- as filtered with our custom dictionary.}
    \textmd{\footnotesize On the left, we observe terminologies pertinent to CTI, such as ``information,'' ``data,'' and ``hacking,'' alongside terms like ``cards,'' which we categorize under cybercrime. Conversely, on the right, we observe more words associated with the selling of tangible goods and their distribution.}} 
    \label{fig:diff}
    \vspace{-1mm}
\end{figure}

\noindent
The final dictionary is curated in a structured approach, which we now illustrate in four separate steps:

\begin{enumerate}
    \item \emph{Definition of Initial Dictionary:} Initially, we define a technical and a non-technical dictionary to identify CTI-relevant data. For the non-technical dictionary, we combine the dictionaries of~\cite{gautam2020hacker, hossen2021generating, adewopo2020exploring} as a set and enrich them with the keywords from Elango et al.~\cite{elango2020redefining}. This consolidation of multiple sources ensures a broad coverage of CTI-related terminology. 
    
    The technical dictionary, on the contrary, is composed of two sub-dictionaries: (i) a regex dictionary consisting of 37 unique regular expressions to identify hashes, email addresses, IP addresses, etc., and (ii) a software dictionary including 2,376 unique names of tools and malware commonly used by adversaries according to the state-of-the-art list of software from MITRE ATT\&CK~\cite{mitre-tools}. We exclude any tools or malware with names shorter than four characters. Note that only 0.018\% of all MITRE ATT\&CK software entries fall into this category.\footnote{During the fine-tuning of the dictionary, delineated below, we observe an increased number of false positives for these short terms.} 
    \item \emph{Fine-tuning of the Dictionary:} We randomly select 500 data items from each of the three datasets (1,500 in total) and label them manually (relevant versus irrelevant to CTI). Two experts first label the data sources independently and then discuss disagreements to obtain consensus. This collaborative approach helps mitigate individual biases and ensures a more reliable ground truth for evaluation. Next, we run the dictionaries across the datasets and refine the keywords of the non-technical and technical dictionary in multiple rounds based on the manual review of the results until we obtain a reasonable F-1 score in the training data, cf. Table~\ref{tab:dict-evaluation}.
    \item \emph{Dictionary Validation}: To validate the dictionaries on unseen test data (``test dataset''), we manually annotated 200 randomly selected data items from each dataset (underground forums, chat channels, and darknet websites) and applied the final dictionary, cf. Table~\ref{tab:dict-evaluation}. This step is crucial for assessing the generalizability of our approach and its performance on diverse and previously not seen data. 
    \item \emph{Review of Included versus Excluded Data Items:} As an additional plausibility check, we examine the word frequencies of data items classified as relevant or irrelevant to CTI. Specifically, we drop all stop words and subtract the normalized word frequencies in one category from those in the other to discern the most distinctive terms, cf. Figure~\ref{fig:diff}. Besides validating the classification approach, the aforementioned analysis also provides insights into the linguistic characteristics of CTI-relevant content, which can potentially lead to future improvements of the dictionary.
   
\end{enumerate}

\begin{table}[h!]
    \centering
    \begin{tabular}{l|c|c}
        \hline
        & \textbf{Train} & \textbf{Test} \\ \hline
        \textbf{Precision} & 0.7808 & 0.7267 \\ \hline
        \textbf{Recall} & 0.9450 & 0.9356 \\ \hline
        \textbf{F1 Score} & 0.8511 & 0.8180 \\ \hline
    \end{tabular}
    \vspace{3mm}
    \caption{\textbf{Dictionary Validation.}
    \textmd{\footnotesize We evaluate the performance of our dictionary based on precision, recall, and F1 score in the train and test dataset. We report the final scores for the train dataset -- after multiple rounds of fine-tuning the dictionary, as delineated above.}} 
    \label{tab:dict-evaluation}
    \vspace{-3mm}
\end{table}

We release the final dictionaries in our repository~\cite{repository}. After filtering for CTI relevance, we keep the following number of data items: 1.6 million for underground forums, 430,000 for chat channels, and 13,000 for darknet websites. The detailed results are presented in Table~\ref{tab:relevant-vs-not}.

\subsection{Background: Topic Modeling}
\label{sapp:topic-modeling}
\noindent
Topic modeling is an unsupervised NLP technique with the main objective of identifying and delineating abstract ``topics'' inherent in a corpus of textual data (here posts, chat messages, or websites). Topic modeling is widely applied for exploratory data analysis~\cite{dieng2020topic, blei2010probabilistic,blei2003latent}. The most well-known topic modeling method is LDA~\cite{blei2003latent}, a generative probabilistic model for textual data. However, LDA considers documents and topics as Bag of Words (BoW), and can hence not account for context, i.e., the relationship between adjacent words in the text. Also, LDA can only include words contained in its training corpus and thus cannot handle unseen words by design~\cite{bianchi2021cross}. Furthermore, the topics inferred by LDA are often of poor quality and misaligned with human judgment~\cite{hinton2009replicated}, as also noted by~\cite{fang2019analyzing} in their analysis of underground forums. 

To overcome the aforementioned challenges, we can use embedding models, enabling a model to account for word context, e.g., realizing that ``Air Canada'' is different than the set of words ``Air'' and ``Canada''~\cite{Mikolov2013}. The combination of pre-trained embeddings with topic modeling led to neural topic models~\cite{li2024improving}. Examples of neural topic models are Contextualized Topic Modeling (CTM)~\cite{bianchi2021cross} or BERTopic~\cite{grootendorst2022bertopic}. However, the core idea of BERTopic and CTM differs substantially. The decisive difference is that BERTopic allows the identification of outliers (i.e., data items that cannot be matched to any topic, and provides additional outlier reduction techniques),\footnote{One exemption is that when K-Means is used for clustering in BERTopic every data item will be assigned to a topic; while when selecting HDBSCAN for clustering (which is the default parameter) outliers are automatically created.} while CTM does not allow outliers by default.

%% file: main_IEEE.bbl
\begin{thebibliography}{10}
\providecommand{\url}[1]{#1}
\csname url@samestyle\endcsname
\providecommand{\newblock}{\relax}
\providecommand{\bibinfo}[2]{#2}
\providecommand{\BIBentrySTDinterwordspacing}{\spaceskip=0pt\relax}
\providecommand{\BIBentryALTinterwordstretchfactor}{4}
\providecommand{\BIBentryALTinterwordspacing}{\spaceskip=\fontdimen2\font plus
\BIBentryALTinterwordstretchfactor\fontdimen3\font minus \fontdimen4\font\relax}
\providecommand{\BIBforeignlanguage}[2]{{%
\expandafter\ifx\csname l@#1\endcsname\relax
\typeout{** WARNING: IEEEtran.bst: No hyphenation pattern has been}%
\typeout{** loaded for the language `#1'. Using the pattern for}%
\typeout{** the default language instead.}%
\else
\language=\csname l@#1\endcsname
\fi
#2}}
\providecommand{\BIBdecl}{\relax}
\BIBdecl

\bibitem{takaaki2019dark}
S.~Takaaki and A.~Inomata, ``Dark web content analysis and visualization,'' in \emph{Proceedings of the ACM International Workshop on Security and Privacy Analytics}.\hskip 1em plus 0.5em minus 0.4em\relax ACM, 2019, pp. 53--59.

\bibitem{bijmans2021catching}
H.~Bijmans, T.~Booij, A.~Schwedersky, A.~Nedgabat, and R.~van Wegberg, ``Catching phishers by their bait: Investigating the dutch phishing landscape through phishing kit detection,'' in \emph{30th USENIX Security Symposium}, 2021, pp. 3757--3774.

\bibitem{sun2019understanding}
Z.~Sun, C.~E. Rubio-Medrano, Z.~Zhao, T.~Bao, A.~Doupé, and G.-J. Ahn, ``Understanding and predicting private interactions in underground forums,'' in \emph{Conference on Data and Application Security and Privacy}.\hskip 1em plus 0.5em minus 0.4em\relax ACM, 2019, pp. 303--314.

\bibitem{carminati2011probability}
B.~Carminati, E.~Ferrari, S.~Morasca, and D.~Taibi, ``A probability-based approach to modeling the risk of unauthorized propagation of information in on-line social networks,'' in \emph{Conference on Data and Application Security and Privacy}.\hskip 1em plus 0.5em minus 0.4em\relax ACM, 2011, pp. 51--62.

\bibitem{jin2011towards}
L.~Jin, H.~Takabi, and J.~B. Joshi, ``Towards active detection of identity clone attacks on online social networks,'' in \emph{Conference on Data and Application Security and Privacy}.\hskip 1em plus 0.5em minus 0.4em\relax ACM, 2011, pp. 27--38.

\bibitem{jeziorowski2020towards}
S.~Jeziorowski, M.~Ismail, and A.~Siraj, ``Towards image-based dark vendor profiling: An analysis of image metadata and image hashing in dark web marketplaces,'' in \emph{International Workshop on Security and Privacy Analytics}.\hskip 1em plus 0.5em minus 0.4em\relax ACM, 2020, pp. 15--22.

\bibitem{saeed2023systematic}
S.~Saeed, S.~A. Suayyid, M.~S. Al-Ghamdi, H.~Al-Muhaisen, and A.~M. Almuhaideb, ``A systematic literature review on cyber threat intelligence for organizational cybersecurity resilience,'' \emph{Sensors}, vol.~23, no.~16, 2023.

\bibitem{islam2021hackerscope}
R.~Islam, M.~O.~F. Rokon, A.~Darki, and M.~Faloutsos, ``Hackerscope: The dynamics of a massive hacker online ecosystem,'' \emph{Social Network Analysis and Mining}, vol.~11, no.~1, 2021.

\bibitem{grootendorst2022bertopic}
M.~Grootendorst, ``Bertopic: Neural topic modeling with a class-based tf-idf procedure,'' \emph{arXiv preprint arXiv:2203.05794}, 2022.

\bibitem{sakellariou2023secdfan}
G.~Sakellariou, P.~Fouliras, and I.~Mavridis, ``Secdfan: A cyber threat intelligence system for discussion forums utilization,'' \emph{Eng}, vol.~4, no.~1, pp. 615--634, 2023.

\bibitem{christin2024learned}
N.~Christin, ``{What I learned from spending a dozen years in the dark web},'' in \emph{Proceedings of the 17th ACM International Conference on Web Search and Data Mining}, 2024, pp. 2--3.

\bibitem{basheer2021threats}
R.~Basheer and B.~Alkhatib, ``Threats from the dark: a review over dark web investigation research for cyber threat intelligence,'' \emph{Journal of Computer Networks and Communications}, vol. 2021, no.~1, 2021.

\bibitem{repository}
``{Our repository},'' \url{https://github.com/irdin-pekaric/WACCO2025/}.

\bibitem{sauerwein2019analysis}
C.~Sauerwein, I.~Pekaric, M.~Felderer, and R.~Breu, ``An analysis and classification of public information security data sources used in research and practice,'' \emph{Computers \& Security}, vol.~82, pp. 140--155, 2019.

\bibitem{shin2021twiti}
H.~Shin, W.~Shim, S.~Kim, S.~Lee, Y.~G. Kang, and Y.~H. Hwang, ``\# twiti: Social listening for threat intelligence,'' in \emph{Proceedings of the Web Conference}, 2021, pp. 92--104.

\bibitem{koloveas2021intime}
P.~Koloveas, T.~Chantzios, S.~Alevizopoulou, S.~Skiadopoulos, and C.~Tryfonopoulos, ``{inTime: A machine learning-based framework for gathering and leveraging web data to cyber-threat intelligence},'' \emph{Electronics}, vol.~10, no.~7, 2021.

\bibitem{tekin2021obtaining}
U.~Tekin and E.~N. Yilmaz, ``Obtaining cyber threat intelligence data from twitter with deep learning methods,'' in \emph{2021 5th International Symposium on Multidisciplinary Studies and Innovative Technologies}.\hskip 1em plus 0.5em minus 0.4em\relax IEEE, 2021, pp. 82--86.

\bibitem{adewopo2020exploring}
V.~Adewopo, B.~Gonen, and F.~Adewopo, ``Exploring open source information for cyber threat intelligence,'' in \emph{IEEE International Conference on Big Data}.\hskip 1em plus 0.5em minus 0.4em\relax IEEE, 2020, pp. 2232--2241.

\bibitem{zhao2020timiner}
J.~Zhao, Q.~Yan, J.~Li, M.~Shao, Z.~He, and B.~Li, ``Timiner: Automatically extracting and analyzing categorized cyber threat intelligence from social data,'' \emph{Computers \& Security}, 2020.

\bibitem{kristiansen2020cti}
L.-M. Kristiansen, V.~Agarwal, K.~Franke, and R.~S. Shah, ``Cti-twitter: Gathering cyber threat intelligence from twitter using integrated supervised and unsupervised learning,'' in \emph{International Conference on Big Data}.\hskip 1em plus 0.5em minus 0.4em\relax IEEE, 2020, pp. 2299--2308.

\bibitem{rodriguez2020enhancing}
A.~Rodriguez and K.~Okamura, ``Enhancing data quality in real-time threat intelligence systems using machine learning,'' \emph{Social Network Analysis and Mining}, vol.~10, no.~1, 2020.

\bibitem{bose2021tracing}
A.~Bose, S.~G. Sundari, V.~Behzadan, and W.~H. Hsu, ``Tracing relevant twitter accounts active in cyber threat intelligence domain by exploiting content and structure of twitter network,'' in \emph{International Conference on Intelligence and Security Informatics}.\hskip 1em plus 0.5em minus 0.4em\relax IEEE, 2021, pp. 1--6.

\bibitem{purba2023extracting}
M.~D. Purba and B.~Chu, ``Extracting actionable cyber threat intelligence from twitter stream,'' in \emph{International Conference on Intelligence and Security Informatics}.\hskip 1em plus 0.5em minus 0.4em\relax IEEE, 2023, pp. 1--6.

\bibitem{sufi2023new}
F.~Sufi, ``A new social media-driven cyber threat intelligence,'' \emph{Electronics}, vol.~12, no.~5, p. 1242, 2023.

\bibitem{sufi2023global}
------, ``A global cyber-threat intelligence system with artificial intelligence and convolutional neural network,'' \emph{Decision Analytics Journal}, vol.~9, 2023.

\bibitem{zenebe2019cyber}
A.~Zenebe, M.~Shumba, A.~Carillo, and S.~Cuenca, ``Cyber threat discovery from dark web,'' \emph{EPiC Series in Computing}, vol.~64, pp. 174--183, 2019.

\bibitem{zhou2022cti}
Y.~Zhou, Y.~Tang, M.~Yi, C.~Xi, and H.~Lu, ``{CTI view: APT threat intelligence analysis system},'' \emph{Security and Communication Networks}, vol. 2022, no.~1, 2022.

\bibitem{panagiotou2021towards}
P.~Panagiotou, C.~Iliou, K.~Apostolou, T.~Tsikrika, S.~Vrochidis, P.~Chatzimisios, and I.~Kompatsiaris, ``Towards selecting informative content for cyber threat intelligence,'' in \emph{International Conference on Cyber Security and Resilience}.\hskip 1em plus 0.5em minus 0.4em\relax IEEE, 2021, pp. 354--359.

\bibitem{sun2021automatic}
T.~Sun, P.~Yang, M.~Li, and S.~Liao, ``An automatic generation approach of the cyber threat intelligence records based on multi-source information fusion,'' \emph{Future Internet}, p.~40, 2021.

\bibitem{ampel2020labeling}
B.~Ampel, S.~Samtani, H.~Zhu, S.~Ullman, and H.~Chen, ``Labeling hacker exploits for proactive cyber threat intelligence: A deep transfer learning approach,'' in \emph{International Conference on Intelligence and Security Informatics}.\hskip 1em plus 0.5em minus 0.4em\relax IEEE, 2020, pp. 1--6.

\bibitem{azevedo2019pure}
R.~Azevedo, I.~Medeiros, and A.~Bessani, ``Pure: Generating quality threat intelligence by clustering and correlating osint,'' in \emph{18th IEEE International Conference On Trust, Security And Privacy In Computing And Communications/13th IEEE International Conference On Big Data Science And Engineering}.\hskip 1em plus 0.5em minus 0.4em\relax IEEE, 2019, pp. 483--490.

\bibitem{guo2023framework}
Y.~Guo, Z.~Liu, C.~Huang, N.~Wang, H.~Min, W.~Guo, and J.~Liu, ``A framework for threat intelligence extraction and fusion,'' \emph{Computers \& Security}, vol. 132, 2023.

\bibitem{cha2020blockchain}
J.~Cha, S.~K. Singh, Y.~Pan, and J.~H. Park, ``Blockchain-based cyber threat intelligence system architecture for sustainable computing,'' \emph{Sustainability}, vol.~12, no.~16, 2020.

\bibitem{furumoto2021extracting}
K.~Furumoto, M.~Umizaki, A.~Fujita, T.~Nagata, T.~Takahashi, and D.~Inoue, ``Extracting threat intelligence related iot botnet from latest dark web data collection,'' in \emph{iThings-GreenCom-CPSCom-SmartData-Cybermatics}.\hskip 1em plus 0.5em minus 0.4em\relax IEEE, 2021, pp. 138--145.

\bibitem{arnold2019dark}
N.~Arnold, M.~Ebrahimi, N.~Zhang, B.~Lazarine, M.~Patton, H.~Chen, and S.~Samtani, ``Dark-net ecosystem cyber-threat intelligence (cti) tool,'' in \emph{International Conference on Intelligence and Security Informatics}.\hskip 1em plus 0.5em minus 0.4em\relax IEEE, 2019, pp. 92--97.

\bibitem{hossen2021generating}
M.~I. Hossen, A.~Islam, F.~Anowar, E.~Ahmed, and M.~M. Rahman, ``Generating cyber threat intelligence to discover potential security threats using classification and topic modeling,'' in \emph{Cyber Security Using Modern Technologies}.\hskip 1em plus 0.5em minus 0.4em\relax CRC Press, 2021, pp. 141--153.

\bibitem{kadoguchi2020deep}
M.~Kadoguchi, H.~Kobayashi, S.~Hayashi, A.~Otsuka, and M.~Hashimoto, ``Deep self-supervised clustering of the dark web for cyber threat intelligence,'' in \emph{International Conference on Intelligence and Security Informatics}.\hskip 1em plus 0.5em minus 0.4em\relax IEEE, 2020, pp. 1--6.

\bibitem{huang2020monitoring}
S.-Y. Huang and T.~Ban, ``Monitoring social media for vulnerability-threat prediction and topic analysis,'' in \emph{International Conference on Trust, Security and Privacy in Computing and Communications}.\hskip 1em plus 0.5em minus 0.4em\relax IEEE, 2020, pp. 1771--1776.

\bibitem{gautam2020hacker}
A.~S. Gautam, Y.~Gahlot, and P.~Kamat, ``Hacker forum exploit and classification for proactive cyber threat intelligence,'' in \emph{Inventive Computation Technologies}.\hskip 1em plus 0.5em minus 0.4em\relax Springer, 2020, pp. 279--285.

\bibitem{arazzi2023nlp}
M.~Arazzi, D.~R. Arikkat, S.~Nicolazzo, A.~Nocera, M.~Conti \emph{et~al.}, ``Nlp-based techniques for cyber threat intelligence,'' \emph{arXiv preprint arXiv:2311.08807}, 2023.

\bibitem{tundis2022feature}
A.~Tundis, S.~Ruppert, and M.~M{\"u}hlh{\"a}user, ``A feature-driven method for automating the assessment of osint cyber threat sources,'' \emph{Computers \& Security}, vol. 113, 2022.

\bibitem{rahman2023attackers}
M.~R. Rahman, R.~M. Hezaveh, and L.~Williams, ``What are the attackers doing now? automating cyberthreat intelligence extraction from text on pace with the changing threat landscape: A survey,'' \emph{ACM Computing Surveys}, vol.~55, no.~12, pp. 1--36, 2023.

\bibitem{chen2019using}
H.~Chen, R.~Liu, N.~Park, and V.~Subrahmanian, ``Using twitter to predict when vulnerabilities will be exploited,'' in \emph{Proceedings of the 25th ACM SIGKDD International Conference on Knowledge Discovery \& Data Mining}, 2019, pp. 3143--3152.

\bibitem{register-chats}
\BIBentryALTinterwordspacing
J.~Burt, ``How cybercrims embrace messaging apps to spread malware, communicate,'' 2022. [Online]. Available: \url{www.theregister.com/2022/08/02/threat_groups_discord_telegram/}
\BIBentrySTDinterwordspacing

\bibitem{threatpost-chats}
\BIBentryALTinterwordspacing
E.~Montalbano, ``Threat actors use telegram to spread ‘eternity’ malware-as-a-service,'' 2022. [Online]. Available: \url{https://threatpost.com/telegram-spread-eternity-maas/179623/}
\BIBentrySTDinterwordspacing

\bibitem{hartel2023darkdiff}
P.~Hartel, E.~Haspels, M.~van Staalduinen, and O.~Texeira, ``Darkdiff: Explainable web page similarity of tor onion sites,'' \emph{arXiv preprint arXiv:2308.12134}, 2023.

\bibitem{Pastrana2018CrimeBBEC}
S.~Pastrana, D.~R. Thomas, A.~Hutchings, and R.~Clayton, ``Crimebb: Enabling cybercrime research on underground forums at scale,'' \emph{World Wide Web Conference}, 2018.

\bibitem{nair2021template}
V.~V. Nair, M.~van Staalduinen, and D.~T. Oosterman, ``Template clustering for the foundational analysis of the dark web,'' in \emph{2021 IEEE International Conference on Big Data}.\hskip 1em plus 0.5em minus 0.4em\relax IEEE, 2021, pp. 2542--2549.

\bibitem{mitre-tools}
``{MITRE ATT\&CK Software},'' \url{https://attack.mitre.org/software/}.

\bibitem{mcinnes2018umap}
L.~McInnes, J.~Healy, and J.~Melville, ``Umap: Uniform manifold approximation and projection for dimension reduction,'' \emph{arXiv preprint arXiv:1802.03426}, 2018.

\bibitem{campello2013density}
R.~J. Campello, D.~Moulavi, and J.~Sander, ``Density-based clustering based on hierarchical density estimates,'' in \emph{Pacific-Asia Conference on Knowledge Discovery and Data Mining}.\hskip 1em plus 0.5em minus 0.4em\relax Springer, 2013.

\bibitem{touvron2023llama}
H.~Touvron, L.~Martin, K.~Stone, P.~Albert, A.~Almahairi, Y.~Babaei, N.~Bashlykov, S.~Batra, P.~Bhargava, S.~Bhosale \emph{et~al.}, ``Llama 2: Open foundation and fine-tuned chat models,'' \emph{arXiv preprint arXiv:2307.09288}, 2023.

\bibitem{abdeen2023smet}
B.~Abdeen, E.~Al-Shaer, A.~Singhal, L.~Khan, and K.~Hamlen, ``Smet: Semantic mapping of cve to att\&ck and its application to cybersecurity,'' in \emph{IFIP Annual Conference on Data and Applications Security and Privacy}.\hskip 1em plus 0.5em minus 0.4em\relax Springer, 2023, pp. 243--260.

\bibitem{pitropakis2019taxonomy}
N.~Pitropakis, E.~Panaousis, T.~Giannetsos, E.~Anastasiadis, and G.~Loukas, ``A taxonomy and survey of attacks against machine learning,'' \emph{Computer Science Review}, vol.~34, 2019.

\bibitem{pekaric2021taxonomy}
I.~Pekaric, C.~Sauerwein, S.~Haselwanter, and M.~Felderer, ``A taxonomy of attack mechanisms in the automotive domain,'' \emph{Computer Standards \& Interfaces}, vol.~78, 2021.

\bibitem{papp2015embedded}
D.~Papp, Z.~Ma, and L.~Buttyan, ``Embedded systems security: Threats, vulnerabilities, and attack taxonomy,'' in \emph{Annual Conference on Privacy, Security and Trust}.\hskip 1em plus 0.5em minus 0.4em\relax IEEE, 2015, pp. 145--152.

\bibitem{pekaric2023systematic}
I.~Pekaric, R.~Groner, T.~Witte, J.~G. Adigun, A.~Raschke, M.~Felderer, and M.~Tichy, ``A systematic review on security and safety of self-adaptive systems,'' \emph{Journal of Systems and Software}, vol. 203, 2023.

\bibitem{groner2023model}
R.~Groner, T.~Witte, A.~Raschke, S.~Hirn, I.~Pekaric, M.~Frick, M.~Tichy, and M.~Felderer, ``Model-based generation of attack-fault trees,'' in \emph{International Conference on Computer Safety, Reliability, and Security}.\hskip 1em plus 0.5em minus 0.4em\relax Springer, 2023, pp. 107--120.

\bibitem{pekaric2024streamlining}
I.~Pekaric, M.~Frick, J.~G. Adigun, R.~Groner, T.~Witte, A.~Raschke, M.~Felderer, and M.~Tichy, ``Streamlining attack tree generation: A fragment-based approach,'' in \emph{57th Annual Hawaii International Conference on System Sciences}, 2024, pp. 7447--7456.

\bibitem{elango2020redefining}
B.~Elango, S.~Matilda, and J.~Jeyasankari, ``Redefining search terms for cybersecurity: A bibliometric perspective,'' in \emph{Proceedings of the International Conference on Recent Advances in Computational Techniques}, 2020.

\bibitem{dieng2020topic}
A.~B. Dieng, F.~J. Ruiz, and D.~M. Blei, ``Topic modeling in embedding spaces,'' \emph{Transactions of the Association for Computational Linguistics}, 2020.

\bibitem{blei2010probabilistic}
D.~Blei, L.~Carin, and D.~Dunson, ``Probabilistic topic models,'' \emph{IEEE Signal Processing Magazine}, vol.~27, no.~6, pp. 55--65, 2010.

\bibitem{blei2003latent}
D.~Blei, A.~Y. Ng, and M.~I. Jordan, ``Latent dirichlet allocation,'' \emph{Journal of Machine Learning Research}, vol.~3, pp. 993--1022, 2003.

\bibitem{bianchi2021cross}
F.~Bianchi, S.~Terragni, D.~Hovy, D.~Nozza, and E.~Fersini, ``Cross-lingual contextualized topic models with zero-shot learning,'' in \emph{16th Conference of the European Chapter of the Association for Computational Linguistics}, 2021.

\bibitem{hinton2009replicated}
G.~E. Hinton and R.~R. Salakhutdinov, ``Replicated softmax: an undirected topic model,'' in \emph{Advances in Neural Information Processing Systems}, vol.~22, 2009.

\bibitem{fang2019analyzing}
Y.~Fang, Y.~Guo, C.~Huang, and L.~Liu, ``Analyzing and identifying data breaches in underground forums,'' \emph{IEEE Access}, 2019.

\bibitem{Mikolov2013}
T.~Mikolov, I.~Sutskever, K.~Chen, G.~S. Corrado, and J.~Dean, ``Distributed representations of words and phrases and their compositionality,'' \emph{Advances in Neural Information Processing Systems}, 2013.

\bibitem{li2024improving}
Z.~Li, A.~Mao, D.~Stephens, P.~Goel, E.~Walpole, A.~Dima, J.~Fung, and J.~Boyd-Graber, ``Improving the tenor of labeling: Re-evaluating topic models for content analysis,'' in \emph{Proceedings of the 18th Conference of the European Chapter of the Association for Computational Linguistics}, 2024, pp. 840--859.

\end{thebibliography}
